%% file: OMClusters_SqueezRes.tex
\documentclass[pra,superscriptaddress,twocolumn
]{revtex4-2}

\usepackage{amsmath,amsfonts,amssymb}
\usepackage{graphicx,bm}
\usepackage{txfonts}
\usepackage{dsfont}
\usepackage{mathrsfs} 
\usepackage{color}
\usepackage{hyperref}

\usepackage[export]{adjustbox}
\usepackage{pstricks}

\newcommand{\rp}[1]{(\ref{#1})}

\newcommand{\abs}[1]{\left|{#1}\right|}

\newcommand{\av}[1]{\left\langle #1 \right\rangle}

\newcommand{\br}[1]{\left\langle #1\right |}
\newcommand{\ke}[1]{|#1\rangle}

\newcommand{\wt}[0]{\widetilde}
\newcommand{\oo}[0]{^{\circ}}
\newcommand{\ooo}[0]{^{\bullet}}
\newcommand{\al}[1]{^{(#1)}}
\newcommand{\da}{^\dagger}

\newcommand{\pt}[1]{\left( #1 \right)}
\newcommand{\pq}[1]{\left[ #1 \right]}
\newcommand{\pg}[1]{\left\{ #1 \right\}}

\newcommand{\lpq}[1]{\left[ #1 \right.}

\newcommand{\rpq}[1]{\left. #1 \right]}

\newcommand{\ee}{{\rm e}}
\newcommand{\ii}{{\rm i}}
\newcommand{\dd}{{\rm d}}

\newcommand{\id}{\openone}

\newcommand{\nn}{{\nonumber}}

\newcommand{\mmat}[2]{
                      \begin{array}{#1}
                       #2
                       \end{array}  }

\newcommand{\matt}[2]{ \pt{
                      \begin{array}{cc}
                       #1 \\
                       #2
                     \end{array}  }  }

\usepackage{mdframed}

\newcommand{\ovl}{\overline}

\newcommand{\va}{{\bf a}}

\newcommand{\ve}{{\bf e}}

\newcommand{\vg}{{\bf g}}

\newcommand{\vv}{{\bf v}}
\newcommand{\vx}{{\bf x}}

\newcommand{\AAA}{{\cal A}}
\newcommand{\BBB}{{\cal B}}
\newcommand{\CC}{{\cal C}}
\newcommand{\DD}{{\cal D}}
\newcommand{\GG}{{\cal G}}
\newcommand{\EE}{{\cal E}}
\newcommand{\FF}{{\cal F}}

\newcommand{\JJ}{{\cal J}}
\newcommand{\KK}{{\cal K}}

\newcommand{\LL}{{\cal L}}
\newcommand{\MM}{{\cal M}}
\newcommand{\NN}{{\cal N}}
\newcommand{\OO}{{\cal O}}

\newcommand{\QQ}{{\cal Q}}
\newcommand{\RR}{{\cal R}}
\newcommand{\TT}{{\cal T}}

\newcommand{\VV}{{\cal V}}
\newcommand{\WW}{{\cal W}}
\newcommand{\XX}{{\cal X}}
\newcommand{\YY}{{\cal Y}}
\newcommand{\ZZ}{{\cal Z}}
\newcommand{\SSS}{{\cal S}}

\definecolor{blue}{rgb}{0,0,0.8}
\definecolor{green}{rgb}{0.,0.6,0.3}
\definecolor{orange}{rgb}{0.9,0.4,0}

\usepackage[normalem]{ulem}

\newcommand{\stkout}[1]{\ifmmode\textrm{\sout{\ensuremath{#1}}}\else\sout{#1}\fi}

\begin{document}

\title{
Multimode Gaussian steady state engineering in optomechanical systems with a squeezed reservoir
}

\author{Nahid Yazdi}
\affiliation{Physics Division, School of Science and Technology, University of Camerino, I-62032 Camerino (MC), Italy}

\author{Stefano Zippilli}
\affiliation{Physics Division, School of Science and Technology, University of Camerino, I-62032 Camerino (MC), Italy}

\author{David Vitali}
\affiliation{Physics Division, School of Science and Technology, University of Camerino, I-62032 Camerino (MC), Italy}
\affiliation{INFN, Sezione di Perugia, I-06123 Perugia, Italy}
\affiliation{CNR-INO, I-50125 Firenze, Italy}

\begin{abstract}
We investigate a theoretical protocol for the dissipative stabilization of mechanical quantum states in a multimode optomechanical system composed of multiple optical and mechanical modes. The scheme employs a single squeezed reservoir that drives one of the optical modes, while the remaining optical modes mediate an effective phonon-phonon interaction Hamiltonian. The interplay between these coherent interactions and the dissipation provided by the squeezed bath enables the steady-state preparation of targeted quantum states of the mechanical modes. In the absence of significant uncontrolled noise sources, the resulting dynamics closely approximate the model introduced in [Phys. Rev. Lett. 126, 020402 (2021)]. We analyze the performance of this protocol in generating mechanical cluster states defined on rectangular graphs.
\end{abstract}

\date{\today}

\maketitle

\section{Introduction}

The precise manipulation and control of quantum states is important for the advancement of quantum technologies. 
Opto- and electromechanical approaches~\cite{barzanjeh2022a,chu2020} are particularly interesting as they allow the quantum control of macroscopic objects and the integration of diverse physical systems~\cite{aspelmeyer2014}.
In particular, multimode opto- and electromechanical systems~\cite{
%
nielsen2017,
weaver2017,
gil-santos2017,
kralj2017,
piergentili2018,
moaddelhaghighi2018,
colombano2019,
mathew2020,
piergentili2021,
fiaschi2021,
mercierdelepinay2021,
mercade2021,
delpino2022,
jong2022,
kharel2022,
youssefi2022,
ren2022a,
mercade2023,
wanjura2023,
madiot2023a,
chegnizadeh2024,
vijayan2024,
alonso-tomas2025,
moores2018,
qiao2023,
vonlupke2024%
}, 
where multiple mechanical modes interact coherently with multiple modes of the electromagnetic field or with superconducting devices, 
offer versatile platforms for the deterministic engineering of complex quantum states~\cite{
ockeloen-korppi2018,
riedinger2018,
kotler2021,
andersson2022,
wollack2022%
}, which may be relevant for various quantum applications.
For example, in quantum communication, such systems can act as quantum repeaters and routers, enabling the faithful transmission and exchange of quantum information among diverse physical platforms~\cite{
stannigel2010,
habraken2012,
asjad2015,
fiaschi2021,
ji2022}; in quantum sensing, quantum features can enhance the sensitivity of measurements~\cite{
ebrahimi2022,
xia2023%
}; and in quantum computing, engineered complex quantum states, can serve as building blocks for quantum computers and simulators~\cite{
stannigelOptomechanicalQuantumInformation2012,
schmidt2012,
ludwig2013,
schmidt2015,
schmidt2015a,
houhou2022,
pechal2018,
hann2019,
chamberland2022%
}. In particular the ability to engineer mechanical cluster states may enable measurement-based quantum computation over mechanical degrees of freedom~\cite{yazdi2024,houhou2022}.

In this work, we describe a protocol for the stabilization of multipartite entangled states of multiple mechanical resonators. 
In this respect, this work is related to Ref.~\cite{yazdi2024}. However, the present protocol relies on distinct dynamics and operates in a different parameter regime. Specifically, we show that a multimode optomechanical system can implement the dissipative model of Ref.~\cite{zippilli2021}, where an array of bosonic modes is driven into tailored Gaussian steady states (including Gaussian cluster states~\cite{zhang2006,zippilli2020}) via the coupling of a single mode to a squeezed reservoir~\cite{clark2017,yap2020}. 
In doing so, this work also generalize the scheme of Ref.~\cite{asjad2016a} to encompass a broader class of steady states.

Here, the bosonic modes considered in Ref.~\cite{zippilli2021} are realized by multiple (quasi-)resonant mechanical resonators, whose interactions are mediated by the optical modes. In this way, we describe a strategy to engineer an effective photon-mediated phonon–phonon interaction Hamiltonian, which enables the implementation of the dissipative preparation scheme of Ref.~\cite{zippilli2021}. As an application, we show how this approach can prepare Gaussian cluster states defined on rectangular graphs.

This article is organized as follows. In Sec.~\ref{Sec:system} we introduce our system. Then in Sec.~\ref{Sec:DissEngineering} we review the main findings of Ref.~\cite{zippilli2021}, and in Sec.~\ref{Sec:approx} we describe how to engineer the model of Ref.~\cite{zippilli2021} with our optomechanical system. 
Numerical results for the preparation of Gaussian cluster states are described in Sec.~\ref{Sec:results}. Finally, we draw our conclusions in Sec.~\ref{conclusions}.
In the Appendix we include additional details on the analytical model (App.~\ref{App:EffectiveModel}), considerations on the preparation of generic cluster states (App.~\ref{App:general-graphs}), description of the evaluation of the system steady state (App.~\ref{App:Standard method}), and of the corresponding fidelity and variance of the nullifiers for the preparation of Gaussian cluster states (App.~\ref{App:fidelity-nullifiers}).

\section{The system}
\label{Sec:system}

\begin{figure}[th!]
\centering
\includegraphics[width=8.5cm]{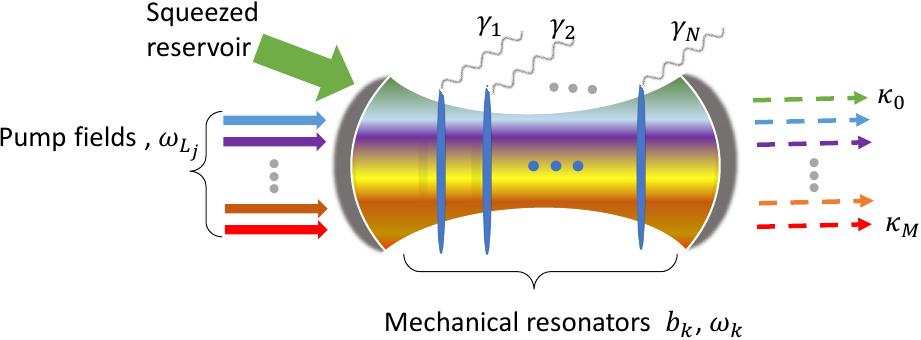}
\caption{
Setup: A multimode optomechanical system where $N$ mechanical resonators interact with $M$ optical modes (here represented by different colors inside the cavity). By controlling the external pump fields it is possible to engineer effective phonon-phonon interactions between the mechanical resonators. The zero-th optical mode is also coupled to a squeezed reservoir. If uncontrolled dissipation is sufficiently weak, the interplay between the dissipation due to the squeezed bath and the coherent photon-mediated interactions is able to stabilize mechanical cluster states.
}
\label{fig0}
\end{figure}

We consider an optomechanical system (see Fig.~\ref{fig0}) where $N$ mechanical modes at frequencies $\omega_k$ interact with $M+1$ optical modes at frequencies $\omega_{cj}$ (the exact value of $M$ depends on the specific state we aim to prepare as specified below).
The mechanical modes are near resonant 
\begin{eqnarray}\label{omegak}
\omega_k&=&\omega_0+\delta\omega_k\ ,
\end{eqnarray}
with $\delta\omega_k\ll\omega_k$, and one optical mode (the one with index
$j=0$) is coupled to a squeezed reservoir~\cite{asjad2016a}. The system works in a regime in which the other optical modes mediate an effective unitary phonon-phonon interaction. In particular, we show that it is to possible engineer the corresponding Hamiltonian to support the creation of a Gaussian cluster state following the procedure discussed in
Ref.~\cite{zippilli2021}.

The optical modes are driven by laser fields at frequencies $\omega_{Lj}$ detuned by $\Delta_j=\omega_{cj}-\omega_{Lj}$ from the cavity resonances. 
The relevant degrees of freedom in quantum optomechanics~\cite{asjad2016a,yazdi2024}  are the quantum fluctuations around the classical steady state. Here they are described by the bosonic annihilation operators $b_k$ and $a_j$, with $k\in\pg{1,...N}$ and $j\in\pg{0,...M}$,
for the mechanical and optical fields respectively. The corresponding optomechanical Hamiltonian in the interaction picture with respect to the optical Hamiltonian is
\begin{eqnarray}
H_{OM}&=&
\hbar\lpq{
\sum_{j=0}^M\ \Delta_j\ a_j\da\,a_j
+
\sum_{k=1}^N\ \omega_k\ b_k\da\,b_k
}
\nn\\&&
\rpq{
+\sum_{j=0}^M\sum_{k=1}^N\ \pt{\GG_{j,k}^*\,a_j+\GG_{j,k}\,a_j\da}\pt{b_k+b_k\da}
}
\end{eqnarray}
where the coupling strengths are proportional to the amplitude of the driving fields
\begin{eqnarray}\label{GGjk}
\GG_{j,k}=g_{jk}\oo\ \alpha_j
\end{eqnarray}
where $g_{jk}\oo$ is the bare optomechanical coupling strength, that depends on system's geometry, optical and mechanical modes structure, and operating point~\cite{biancofiore2011, karuza2013, eichenfield2009, wu2014a, shevchuk2017, rodrigues2019}, and $\alpha_j$ is the classical complex steady state amplitude of the $j$-th optical mode, that can be controlled through the driving lasers amplitudes $E_j$ and phases $\phi_{Lj}$ according to the relation  
\begin{eqnarray}
\alpha_j=\frac{-\ii\, E_j\ \ee^{\ii\,\phi_{L,j}}}{\kappa_j+\ii\,\Delta_j}\ .
\end{eqnarray}
Including optical and mechanical dissipation with rates $\kappa_j$ and $\gamma_k$ respectively, we describe the system dynamics, as customary in quantum optomechanics~\cite{bowen2015}, in terms of the quantum Langevin equations 
\begin{eqnarray}\label{QLE00}
\dot a_j&=&-\pt{\kappa_j+\ii\,\Delta_j}\,a_j-\ii\sum_{k=1}^N\,\GG_{j,k}\pt{b_k+b_k\da}+\sqrt{2\,\kappa_j}\ a_j\al{in}
\nn\\
\dot b_k&=&-\pt{\gamma_k+\ii\,\omega_k}\,b_k-\ii\sum_{j=0}^M\pt{\GG_{j,k}^*\,a_j+\GG_{j,k}\,a_j\da}+\sqrt{2\,\gamma_k}\ b_k\al{in}.
\nn\\
\end{eqnarray}
Differently from the standard quantum optomechanical setting~\cite{yazdi2024}, here the reservoir acting on the zero-th optical mode is squeezed. In practice, this can be achieved by driving the optical mode with the output field of an optical parametric oscillator~\cite{clark2017,yap2020} as detailed, for example, in Ref.~\cite{asjad2016a}.
We describe the squeezing reservoir in terms of the correlations of the corresponding input noise operator $a_0\al{in}$, which, assuming a sufficiently large squeezing bandwidth, are characterized by the relations
\begin{eqnarray}\label{corrSqueezBath}
\av{a_0\al{in}(t)\ a_0\al{in}{}\da(t')}&=&\delta(t-t')+\av{a_0\al{in}{}\da(t)\ a_0\al{in}(t')}
\nn\\&=&
(1+n_s)\ \delta(t-t')
\nn\\
\av{a_0\al{in}(t)\ a_0\al{in}(t')}&=&\av{a_0\al{in}{}\da(t)\ a_0\al{in}{}\da(t')}^*
\nn\\&=&
-\ee^{-2\ii\ \epsilon_{L0} t}
m_s\ \delta(t-t')\ ,
\end{eqnarray}
where $\epsilon_{L0}=\omega_s-\omega_{L0}$ is the detuning between the central frequency $\omega_s$ of the squeezed reservoir and the frequency of the driving laser of the zero-th optical field~\cite{asjad2016a}, $n_s=\sinh(r)^2$, and $m_s=\ee^{\ii\ \varphi_0}\ \sinh(r)\ \cosh(r)$, with $r$ the squeezing parameter of the reservoir and $\varphi_0$ the squeezing phase.
The other non-zero correlation functions involving all the remaining input noise operators are
\begin{eqnarray}\label{mechnoisecorrelations}
\av{a_j\al{in}(t)\ a_{j}\al{in}{}\da(t')}&=& \delta(t-t')
\nn\\
\av{b_k\al{in}(t)\ b_{k}\al{in}{}\da(t')}&=&\delta(t-t')+\av{b_k\al{in}{}\da(t)\ b_{k}\al{in}(t')}
\nn\\&=&
(1+\ovl n_{T,k})\ \delta(t-t')\ ,
\end{eqnarray}
where $\ovl n_{T,k}=\pt{\ee^{\hbar\,\omega_k/k_BT}-1}^{-1}$ is the average number of thermal excitations corresponding to the $k$-th mechanical mode.

Here we aim at engineering a model analogous to the one analyzed in Ref.~\cite{zippilli2021}. The modes that we expect to entangle in the steady state are the mechanical ones and their Hamiltonian is realized by the optical fields that mediate the interactions between the mechanical modes.

\section{Model and results of Ref.~\cite{zippilli2021}}
\label{Sec:DissEngineering}

Let us now briefly review and rephrase the results of Ref.~\cite{zippilli2021}.

Ref.~\cite{zippilli2021} analyzes an array of $N+1$ bosonic modes. Here we indicate the corresponding annihilation operators with the symbols $c_\ell$ for $\ell\in\pg{0,...N}$. In our system they correspond to the zero-th optical mode and the $N$ mechanical modes (see Sec.\ref{Sec:effModel} below). 
The modes interact according to a passive Hamiltonian 
\begin{eqnarray}\label{H}
H=\hbar\sum_{\ell,\ell'=0}^N\ \JJ_{\ell,\ell'}\ c_\ell\da\ c_{\ell'}\ ,
\end{eqnarray}
with $\JJ$ an $(N+1)\times(N+1)$ hermitian matrix and, in the ideal case, only the zero-th mode is dissipative and coupled, with rate $\kappa_0$, to a squeezed reservoir, such that the system dynamics can be described by the quantum Langevin equations
\begin{eqnarray}\label{QLEideal}
\dot c_0
&=&
-\kappa_0\ c_0-\ii\sum_{k=0}^N\,\JJ_{0,k}\ c_k
+\sqrt{2\,\kappa_0}\ c_0\al{in}
\nn\\
\dot c_k
&=&
-\ii\sum_{k'=0}^N \JJ_{k,k'}\ c_{k'}
\ ,
\end{eqnarray}
where the input noise operator $c_0\al{in}$ fulfill the correlation relations for a squeezed reservoir similar to Eq.~\rp{corrSqueezBath} but without the time dependent phase,
\begin{eqnarray}\label{Corrc0}
\av{c_0\al{in}(t)\ c_0\al{in}{}\da(t')}
&=&
\delta(t-t')+\av{c_0\al{in}{}\da(t)\ c_0\al{in}(t')}
\nn\\&=&
(1+n_s)\ \delta(t-t')
\nn\\
\av{c_0\al{in}(t)\ c_0\al{in}(t')}
&=&
\av{c_0\al{in}{}\da(t)\ c_0\al{in}{}\da(t')}^*
\nn\\&=&
-m_s\ \delta(t-t')\ .
\end{eqnarray} 
Ref.~\cite{zippilli2021} shows that such system 
can have, as unique pure steady state, a specific Gaussian state obtained by applying a passive unitary transformation to a set of equally squeezed mode. Specifically, the steady state can take the form 
\begin{eqnarray}\label{Uket0}
\ke{\Psi_{\rm tot}}&=&
\pt{U\al{p}U\al{S}\ke{0}}\ \otimes\ \pt{U_0\ke{0}}\ ,
\end{eqnarray}
with $U\al{S}$ and $U_0$ the squeezing transformations for all the modes
\begin{eqnarray}\label{U0US}
U_0&=&\ee^{\frac{z}{2}\pt{\ee^{\ii\varphi_0}\ c_0\da{}^2-\ee^{-\ii\varphi_0}\ c_0^2}}
\nn\\
U\al{S}&=&\prod_{k=1}^N
\ee^{\frac{z}{2}\pt{\ee^{\ii\varphi_k}\ c_k\da{}^2-\ee^{-\ii\varphi_k}\ c_k^2}}
\end{eqnarray}
and $U\al{p}$ a passive unitary for the modes not directly coupled to the reservoir
\begin{eqnarray}\label{Up}
U\al{p}&=&\ee^{-\ii\sum_{k,k'=1}^N\ \KK_{k,k'}\ c_k\da\ c_{k'}}\ ,
\end{eqnarray}
with $\KK$ an $N\times N$ hermitian matrix. 

\subsection{
Sufficient condition for a pure steady state of the form of Eq.~\rp{Uket0}
}

A sufficient condition for the model in Eq.~\rp{QLEideal} to admit a unique pure steady state of the form given by Eqs.~\rp{Uket0}-\rp{Up} is that 
the system Hamiltonian is passive, as specified by Eq.~\rp{H},
and it can be expressed as
\begin{eqnarray}\label{HUHU}
H=U\al{p}\ H\al{S}\ U\al{p}{}\da\ ,
\end{eqnarray}
where $H\al{S}$ is the Hamiltonian for a linear chain with nearest-neighbor couplings
\begin{eqnarray}\label{HS}
H\al{S}&=&
\ii\,\hbar\sum_{\ell,\ell'=0}^N\ 
\ee^{\ii\,\frac{\varphi_{\ell}-\varphi_{\ell'}}{2}}\ 
\JJ_{\ell,\ell'}\al{S}\ c_\ell\da\ c_{\ell'}
\ ,
\end{eqnarray}
where
$\JJ\al{S}$ 
is the $(N+1)\times(N+1)$ real anti-symmetric matrix
\begin{eqnarray}\label{JJS}
\JJ\al{S}&=&
\matt{0       & -J_0\ \ve^T}
{J_0\ \ve & \ovl\JJ\al{S}}\ ,
\end{eqnarray}
and where for later convenience we introduced the vector 
\begin{eqnarray}
\ve=\pt{\mmat{c}{
1\\0\\ \vdots\\0}}\ ,
\end{eqnarray}
and the $N\times N$ matrix
\begin{eqnarray}\label{ovlJJS}
\ovl\JJ\al{S}&=&\pt{
\mmat{ccccc}{
0      & -J_1   & 0      & \cdots  & 0        \\
J_1    & 0      & -J_2   &         &          \\
0      &  J_2   & \ddots & \ddots  &          \\
\vdots &        & \ddots &         & -J_{N-1} \\
0      &        &        & J_{N-1} & 0 
}}\ .
\end{eqnarray}
This specific form of the system Hamiltonian~\rp{HUHU} guarantees a unique steady state. Indeed, as proved in Ref.~\cite{zippilli2021}, 
in this way all the normal modes of the array are coupled to the zeroth dissipative mode, so that all normal modes can dissipate and the steady state is unique.

This can be understood by analyzing the system dynamics in terms of the transformed operators
\begin{eqnarray}\label{wtck}
\wt c_j=U\al{p}{}\da\ U\al{S}{}\da\ U_0\da\ c_j\ U\al{p}\ U\al{S}\ U_0\ .
\end{eqnarray}
An important property of $H\al{S}$~\rp{HS} is that it is invariant under the squeezing transformations~\rp{U0US}~\cite{zippilli2021}.
As a consequence, using also Eq.~\rp{HUHU}, we
find that the transformed modes satisfy the equations
\begin{eqnarray}\label{QLEsqueezed}
\dot{\wt c}_0
&=&
-\kappa_0\ \wt c_0+\sum_{k=0}^N\,
\ee^{\ii\,\frac{\varphi_0-\varphi_k}{2}}\
\JJ_{0,k}\al{S}\ \wt c_k
+\sqrt{2\,\kappa_0}\ \wt c_0\al{in}
\nn\\
\dot{\wt c}_k
&=&
\sum_{k'=0}^N 
\ee^{\ii\,\frac{\varphi_k-\varphi_k'}{2}}\
\JJ_{k,k'}\al{S}\ \wt c_{k'}
\ ,
\end{eqnarray}
where $\wt c_0\al{in}=\cosh(r)\ c_0\al{in}+\ee^{\ii\,\varphi_0}\ \sinh(r)\ c_0\al{in}{}\da$ describes vacuum noise, such that the only non-zero correlation function is $\av{c_0\al{in}(t)\ c_0\al{in}{}\da(t')}=\delta(t-t')$. These equations describe the dynamics of the linear chain~\rp{HS}-\rp{ovlJJS} with an open end that dissipates into a vacuum reservoir. Therefore, these modes lose excitations during the dynamics and approach the vacuum $\ke{0}$ in the steady state. In fact, all the normal modes of a linear chain (whose annihilation operators can be expressed as the linear combinations $\sum_{k'}\ \pg{\vv\al{k}}_{k'}\ \wt c_{k'}$, where $\vv\al{k}$ are eigenvectors of $\JJ\al{S}$ satisfying $\JJ\al{S}\ \vv\al{k}=\lambda_k\ \vv\al{k}$) have a finite overlap with the end mode (that is, $\pg{\vv\al{k}}_{0}\neq 0$ for all $k$). Hence, all normal modes are coupled to the reservoir and can dissipate through the open end mode $\wt c_0$. 
This means that the system has no dark modes and that the steady state is unique. 
Correspondingly, since $H\al{S}$ is invariant under $U\al{S}\,U_0$ and $U\al{p}$ does not operate on the zeroth mode, all the normal modes of $H$ also have a finite overlap with the original zeroth mode $c_0$. Hence, the original system has no dark modes, and its steady state is unique. Specifically, the steady state of the original modes $c_k$ is obtained by transforming the steady state for the transformed modes~$\wt c_k$ under the unitary $U\al{P}\ U\al{S} U_0$ that relates these modes [see Eq.~\rp{wtck}], namely, it is equal to Eq.~\rp{Uket0}.

In Sec.~\ref{Sec:approx} we show that our optomechanical system can be engineered to approximate the ideal model~\rp{QLEideal}, up to small uncontrolled noise, with a Hamiltonian of the form~\rp{HUHU}-\rp{ovlJJS}. As a consequence, the steady state of the optomechanical system is expected to closely approximate the pure state defined in Eqs.~\rp{Uket0}-\rp{Up}. 

\subsection{
A convenient rewriting of the system Hamiltonian
}

For the purpose of this work it is useful to rewrite the passive system Hamiltonian~\rp{HUHU}-\rp{ovlJJS}, that guarantees a unique pure steady state~\rp{Uket0},  as follows.  
We consider the matrix of coefficient $\JJ$ 
[see Eq.~\rp{H}] and we express it as the block matrix
\begin{eqnarray}\label{JJ1}
\JJ&=&\pt{
\mmat{ccccc}{
0      & \vg_0\al{\JJ}{}^T\\
\vg_0\al{\JJ}{}^*    & \WW\al{\JJ}
}}\ ,
\end{eqnarray}
(the symbol ${}^*$ indicates the element wise complex conjugate) where the blocks $\vg_0\al{\JJ}$ and $\WW\al{\JJ}$ can be expressed in terms of the $N\times N$ unitary matrix 
\begin{eqnarray}\label{VV}
\VV=\ee^{-\ii\,\KK}\ ,
\end{eqnarray}
where $\KK$ determines $U\al{p}$ according to Eq.~\rp{Up}. In fact, $\VV$ defines the transformation of the operators $c_k$, under the effect of $U\al{p}$ itself, i.e.
\begin{eqnarray}
U\al{p}\ c_k\ U\al{p}{}\da&=&\sum_{k'=1}^N\ \pg{\VV\da}_{k,k'}\ c_{k'}\ ,
\hspace{0.3cm}{\rm for}\ k\in\pg{1,...N}\ .
\end{eqnarray}
And so, considering also that 
$U\al{p}\ c_0\ U\al{p}{}\da=c_0$, the Hamiltonian~\rp{HUHU}-\rp{ovlJJS}
can be rewritten as 
\begin{eqnarray}\label{H(V)}
H&=&\ii\,\hbar\ J_0\al{S}\pq{
\ee^{\ii\,\frac{\varphi_{1}-\varphi_0}{2}}\ 
\sum_{k=1}^N\ \VV_{k,1}\ 
c_{k}\da\ c_0
-h.c.
}
\nn\\&&
+\ii\,\hbar\sum_{k,k'=1}^N\
\pg{
\VV\ \Phi\ \ovl\JJ\al{S}\ \Phi\da\ \VV\da
}_{k,k'}\ c_k\da\ c_{k'}\ ,
\end{eqnarray}
where $\Phi$ is the diagonal $N\times N$ matrix with non-zero elements $\Phi_{k,k}=\ee^{\ii\,\frac{\varphi_k}{2}}$.
As a result, $\WW\al{\JJ}$ in Eq.~\rp{JJ1}, is given by
\begin{eqnarray}\label{WWJ}
\WW\al{\JJ}=\ii\ \VV\ \Phi\ \ovl\JJ\al{S}\ \Phi^*\ \VV\da\ ,
\end{eqnarray}
and $\vg_0\al{\JJ}$ in Eq.~\rp{JJ1}, is the vector
\begin{eqnarray}\label{vg0J}
\vg_0\al{\JJ}=-\ii\ J_0\ \ee^{-\ii\,\frac{\varphi_{1}-\varphi_0}{2}}\ \VV^*\,\ve\ .
\end{eqnarray}

This means that given a target state of the form~\rp{Uket0}, which is fully characterized by $\VV$~\rp{VV} (i.e. $\KK$) and $\Phi$ for any value of the squeezing parameter $z$, we can identify the Hamiltonian~\rp{H} [with matrix of coefficient~\rp{JJ1}], that support the given steady state, directly using Eqs.~\rp{WWJ} and \rp{vg0J}. In particular, since $U\al{p}$ is arbitrary, this strategy allows one to steer the $N$ modes not directly coupled to the squeezed reservoir into any zero-average Gaussian pure state
\begin{eqnarray}\label{Psi}
\ke{\Psi}=U\al{p}U\al{S}\ke{0}\ ,
\end{eqnarray} 
that can be obtained by applying any passive transformation $U\al{p}$ over many equally squeezed modes $U\al{S}\ke{0}$.

\subsection{Application to the steady state preparation of a cluster state}
\label{Sec:ClusterState0}

Ref.~\cite{zippilli2021} describes also the application of this general result to the preparation of a Gaussian cluster state~\cite{zhang2006,zippilli2020}. 

An ideal Gaussian cluster state with adjacency matrix $\AAA$ (symmetric, with nonzero elements equal to one and $\AAA_{k,k}=0$) is defined as the eigenstate with zero eigenvalue of the nullifiers, which are linear combinations of quadrature operators~\cite{zhang2006,zippilli2020}
\begin{equation}\label{nulli}
X_k = -\ii\!\left(b_k e^{\ii\theta_k}-b_k^{\dagger}e^{-\ii\theta_{k}}\right) 
- \sum_{k'=1}^{N}\mathcal{A}_{kk'}\!\left(b_{k'} e^{\ii\theta_{k'}}+b_{k'}^{\dagger}e^{-\ii\theta_{k'}}\right).
\end{equation}
This means that an ideal cluster state $\lvert \Psi_{\rm cluster}\al{\rm ideal} \rangle$ is determined by the conditions
\begin{equation}
	X_k \, \lvert \Psi_{\rm cluster}\al{\rm ideal} \rangle = 0 , \qquad \forall k.
\end{equation}
As a result, it is infinitely squeezed.
A realistic cluster state (as the one introduced hereafter) has finite squeezing, and the nullifiers exhibit finite variances.
Cluster states are an essential resource for measurement-based quantum computation, where a computation is performed as a sequence of measurements on this kind of state~\cite{menicucci2006,gu2009}.

According to Ref.~\cite{zippilli2020}, an approximated Gaussian cluster state, $\ke{\Psi_{\rm cluster}}$, with adjacency matrix $\AAA$ can be generated from the vacuum  by a multimode squeezing transformation 
\begin{eqnarray}\label{U}
U=\ee^{\ii\frac{z}{2}\sum_{k,k'=1}^N\pt{\ZZ_{k,k'} c_k\da\ c_{k'}\da+h.c. } }\, 
\end{eqnarray}
i.e. $\ke{\Psi_{\rm cluster}}=U\ke{0}$,
when the matrix of coefficients $\ZZ$ is given by
\begin{eqnarray}\label{ZZ}
\ZZ=\Theta\ \ZZ_0\ \Theta\ ,
\end{eqnarray}
where
\begin{eqnarray}\label{ZZ0}
\ZZ_0=-\ii\ \frac{\AAA-\ii\ \id}{\AAA+\ii\ \id}\ ,
\end{eqnarray}
and $\Theta$ is a diagonal matrix with elements $\Theta_{k,k}=\ee^{-\ii\,\theta_k}$, such that $\ZZ$ is a symmetric unitary matrix. 
In fact, more in general, the state $\ke{\Psi_{\rm cluster}}$ is equivalently generated by $U U\al{x}$, where $U\al{x}=
\ee^{-\ii\sum_{k,k'=1}^N\ \KK_{k,k'}\al{x}\ c_k\da\ c_{k'}}
$, with $\KK\al{x}$ hermitian, is a generic passive unitary that have no effect on the vacuum $U\al{x}\ke{0}=\ke{0}$, namely $\ke{\Psi_{\rm cluster}}=U\,U\al{x}\ke{0}$.
Furthermore, Ref.~\cite{zippilli2020} also shows that, in this case, the state $\ke{\Psi_{\rm cluster}}$  can be expressed in terms of a multimode passive transformation $U\al{p}$ applied to many equally squeezed modes $U\al{S}\ke{0}$ such that
\begin{eqnarray}\label{Psicluster}
\ke{\Psi_{\rm cluster}}&=&
U\ U\al{x}\ke{0}
\nn\\&=&
U\al{p}\ U\al{S}\ke{0}
\end{eqnarray}
[see Eq.~\rp{Psi}]. And this entails that it can be generated with the dissipative model of Ref.~\cite{zippilli2021}. In order to apply the results of the previous section to the generation of the Gaussian cluster state~\rp{Psicluster}, we need to determine a relation between the matrix $\ZZ$~\rp{ZZ}, that determines both the cluster state and the unitary $U$~\rp{U}, and the matrix $\VV$~\rp{VV}, that determines the unitary $U\al{p}$~\rp{Up}.
This can be achieved by considering how the mode operator $c_k$ transforms under both $U\,U\al{x}$ and $U\al{p}\,U\al{S}$. Then, by equating the two transformation relations, we obtain an equation that expresses the relation between $\VV$ and $\ZZ$.
Specifically, introducing $\VV\al{x}=\ee^{-\ii\,\KK\al{x}}$, we find~\cite{zippilli2020,zippilli2021} 
\begin{eqnarray}
U\al{x}{}\da\ U\da\ c_k\ U\ U\al{x}&=&
\nn\\&&\hspace{-3.3cm}
=\sum_{k'=1}^N\pt{
\cosh(z)\ \VV_{k,k'}\al{x}\ c_{k'} -\ii\,\sinh(z)
\pg{\ZZ\ \VV\al{x}{}^*}_{k,k'}\ c_{k'}\da
}
\end{eqnarray}
and 
\begin{eqnarray}\label{UdacU}
U\al{S}{}\da\ U\al{p}{}\da\ c_k\ U\al{p}\ U\al{S}&=&
\nn\\&&\hspace{-3.3cm}
=\sum_{k'=1}^N\pt{
\cosh(z)\ \VV_{k,k'}\ c_{k'} +\sinh(z) \pg{\VV\ \Phi^2}_{k,k'}\ c_{k'}\da
}\ . 
\end{eqnarray}
These two expressions are equal (meaning that $U\,U\al{x}=U\al{p}\,U\al{S}$, so that $U\ke{0}=U\al{p}\,U\al{S}\ke{0}$) when $\VV\al{x}=\VV$ and
\begin{eqnarray}\label{VVPhiVV}
\VV\ \Phi^2\ \VV^T&=&
-\ii\,\Theta\ \ZZ_0\ \Theta\ ,
\end{eqnarray}
that is, when 
\begin{eqnarray}\label{VVZOPhi}
\VV&=&\Theta\sqrt{-\ii\,\ZZ_0}\ \OO_0\ \Phi^*\ ,
\end{eqnarray}
where $\OO_0$ is a generic orthogonal matrix.

In this way, given the adjacency matrix $\AAA$ and $\Theta$, that determine the cluster state through Eqs.~\rp{U}-\rp{ZZ0}, we can determine $\VV$~\rp{VVZOPhi} and correspondingly, using Eqs.~\rp{JJ1}, \rp{WWJ} and \rp{vg0J}, the Hamiltonian~\rp{H} that support the preparation of this state. Here the same cluster state can be prepared for any orthogonal matrix $\OO_0$, for any set of phases that determine $\Phi$, and for any (non-zero real) value of the entries of the matrix $\JJ\al{S}$.

\section{Approximating the model of Ref.~\cite{zippilli2021} with our optomechanical system}
\label{Sec:approx}

In this work, we want to implement the model of Ref.~\cite{zippilli2021} with our optomechanical system. 
Our goal is to control the quantum state of the mechanical modes, such that the state~\rp{Psi} introduced in the previous section is realized for the set of mechanical modes.

In our system, the mechanical modes do not couple directly; instead, their interaction is mediated by the optical modes. 
In the limit of rapidly evolving optical fields, the optical degrees of freedom can be adiabatically eliminated, yielding a reduced model similar to the one of Ref.~\cite{zippilli2021}, i.e. equal to Eq.~\rp{QLEideal}, but with additional noise affecting all the modes. 
This approximate model can therefore exhibit a steady state that closely approximates a state of the form given in Eq.~\rp{Uket0} when
\begin{itemize}
\item[(i)] 
the Hamiltonian of the approximate model has the form given in Eqs.~\rp{HUHU}-\rp{ovlJJS}, or equivalently Eq.~\rp{H(V)} (see Sec.~\ref{Sec:necessaryConditions});

\item[(ii)] 
the additional mechanical noise is sufficiently small (see Sec.~\ref{Sec:validity}).
\end{itemize}

On the one hand, the required effective mechanical Hamiltonian can be engineered by tailoring the optically mediated phonon–phonon interactions. On the other, weak mechanical dissipation (compared with the engineered Hamiltonian terms and with the dissipation in the squeezed reservoir) can be achieved when the intrinsic mechanical damping rate is small and the optical detuning is large, such that the optomechanical processes operate in the dispersive regime, where optically induced mechanical dissipation becomes negligible.

In what follows, after introducing the approximate model, we first outline how to choose the optomechanical interaction strengths $\GG_{jk}$ such that the resulting photon-mediated Hamiltonian coincides with Eq.~\rp{H(V)}. We then analyze the conditions under which optically mediated mechanical dissipation is smaller than the Hamiltonian terms. Finally, we describe the coefficients of the Hamiltonian~\rp{H(V)} which are needed for the preparation of a rectangular Gaussian cluster state.  

\subsection{The effective optomechanical model}\label{Sec:effModel}

When the dynamics of the optical fields is much faster than that of the mechanical degrees of freedom it is possible to adiabatically eliminate the optical degrees of freedom and approximate the mechanical dynamics in terms of an effective model for the mechanical modes that is 
equal to Eq.~\rp{QLEideal}, but with additional noise terms.
To be specific, under the assumption of fast optical dynamics, that corresponds to 
\begin{eqnarray}\label{cond_AdEl}
\abs{\kappa_j+\ii\,\Delta_j}\gg\GG_{j,k}\ ,
\end{eqnarray}
we eliminate the optical modes with indices $j\neq0$, and we approximate the system dynamics in terms of an optomechanical model where a single optical mode (the one with index $j=0$) interacts with many mechanical modes that, in turn, interact according to an effective optic-mediated Hamiltonian that we introduce hereafter.

We establish the equivalence between the array operators $c_k$ introduced in Sec.~\ref{Sec:DissEngineering} (that corresponds to the model of Ref.~\cite{zippilli2021}), and the slowly varying operators
\begin{eqnarray}\label{slowOp}
a_0\ooo(t)&=&a_0(t)\ \ee^{\ii\ \epsilon_{L0}\ t}
\nn\\
b_k\ooo(t)&=&b_k(t)\ \ee^{\ii\ \omega_0\ t}\ ,
\end{eqnarray}
according to
\begin{eqnarray}\label{cb}
c_0&\equiv&a_0\ooo
\nn\\
c_k&\equiv&b_k\ooo\ .
\end{eqnarray}

As discussed in details in App.~\ref{App:EffectiveModel} the effective model, for the slowly varying operators~\rp{slowOp}, with $\epsilon_{L0}=\omega_0=\Delta_0$ (indicating that the central frequency of the squeezed reservoir is resonant with the zero-th cavity mode~\cite{asjad2016a} which, in turn, is driven on the red mechanical sideband), is given by 
a set of quantum Langevin equations of the form
\begin{eqnarray}\label{QLE01}
\dot a_0\ooo&=&-
\kappa_0\ a_0\ooo-\ii\sum_{k=1}^N\,\GG_{0,k}\ b_k\ooo
+\sqrt{2\,\kappa_0}\ a_0\ooo{}\al{in}
\nn\\
\dot b_k\ooo
&=&
-\sum_{k'=1}^N \pt{\YY_{k,k'}+\ii\,\WW_{k,k'}}b_{k'}\ooo
-\ii\,\GG_{0,k}^*\,a_0\ooo
+ \sqrt{2\ \gamma_k}\,b_k\ooo{}\al{in}+y_k\ooo\ ,
\nn\\
\end{eqnarray}
which coincides with Eq.~\rp{QLEideal} when the mechanical noise terms are zero, i.e. $\YY_{k,k'}=\gamma_k=0$ and $y_k\ooo=0$, for all $k,k'$.
We now define all the quantities introduced in these equations.

The correlations of the slowly varying optical input noise 
$a_0\ooo{}\al{in}$
are equal to
those for $c_0\al{in}$ in Eq.~\rp{Corrc0}.

The coherent photon mediated mechanical interactions are described by 
\begin{eqnarray}\label{WW}
\WW_{k,k'}
&=&
\delta\omega_k\ \delta_{k,k'}+\sum_{j=1}^M\ \GG_{j,k}^*\ \GG_{j,k'}\ \DD_{j,j}\ ,
\end{eqnarray}
with $\delta_{k,k'}$ the Kronecker delta and 
$\DD$ the $M\times M$ diagonal matrix with elements 
\begin{eqnarray}\label{DD}
\DD_{j,j}=-\frac{
2\ \Delta_j 
\pt{\kappa_j^2+\Delta_j^2-\omega_0^2}
}{
\pt{\kappa_j^2+\Delta_j^2-\omega_0^2}^2+4\ \kappa_j^2\ \omega_0^2
}\ .
\end{eqnarray}
Therefore, the corresponding effective system Hamiltonian, that we aim to engineer,  can be expressed as 
\begin{eqnarray}\label{Heff}
H\al{eff}&=&
\hbar\sum_{\ell=1}^N\ \pq{
\JJ_{\ell,0}\al{eff}\ b_\ell\ooo{}\da\ a_{0}\ooo
+
\JJ_{0,\ell}\al{eff}\ a_{0}\ooo{}\da\ b_\ell\ooo 
}
\nn\\&&
+\hbar\sum_{\ell,\ell'=1}^N\ \JJ_{\ell,\ell'}\al{eff}\ b_\ell\ooo{}\da\ b_{\ell'}\ooo\ ,
\end{eqnarray}
with $\JJ\al{eff}$ the $\pt{N+1}\times\pt{N+1}$ matrix of coefficients
\begin{eqnarray}\label{JJ2}
\JJ\al{eff}&=&\matt{0 & \vg_0^T}{\vg_0^* & \WW}\ ,
\end{eqnarray}
where we introduced the vector of interaction coefficients $\vg_0$, the element of which are
\begin{eqnarray}\label{vg0}
\pg{\vg_0}_k=\GG_{0,k}\ .
\end{eqnarray}
We rewrite $\WW$~\rp{WW} by introducing the reduced $M\times N$ coupling matrix $\ovl\GG$ which excludes the zero-th optical mode, and the diagonal matrix $\WW\al{\delta}$, with elements 
\begin{eqnarray}\label{Wdelta}
\WW\al{\delta}_{k,k}=\delta\omega_k\ ,
\end{eqnarray}
as
\begin{eqnarray}\label{WW2}
\WW=\WW\al{\delta}+\ovl\GG\da\ \DD\ \ovl\GG\ .
\end{eqnarray}

Moreover, the effective mechanical dissipation (correlated dissipation involving various mechanical modes, or in other terms dissipative coupling)  is described by the matrix of coefficients 
\begin{eqnarray}\label{YY}
\YY_{k,k'}&=&
\gamma_k\ \delta_{k,k'}
+\sum_{j=1}^M
\frac{
4\ \GG_{j,k}^*\ \GG_{j,k'}\ \Delta_j\  \kappa_j\ \omega_0
}{
\pt{\kappa_j^2+\Delta_j^2-\omega_0^2}^2+4\ \kappa_j^2\ \omega_0^2
}\ .
\end{eqnarray}

Finally, the photon-mediated noise operators $y_k\ooo(t)$ are characterized by the correlation functions
\begin{eqnarray}\label{avyy}
\av{y_k\ooo(t)\ y_{k'}\ooo{}\da(t')}&=&2\ \delta(t-t')
\nn\\&&\times
\sum_{j=1}^M \kappa_j\ \GG_{j,k}^*\ \GG_{j,k'}\ \abs{
\frac{1}{\kappa_j+\ii\pt{\Delta_j-\omega_0}}
}^2
\nn\\
\av{y_k\ooo{}\da(t)\ y_{k'}\ooo(t')}&=&2\ \delta(t-t')
\nn\\&&\times
\sum_{j=1}^M \kappa_j\ \GG_{j,k}^*\ \GG_{j,k'}\ \abs{
\frac{1}{\kappa_j+\ii\pt{\Delta_j+\omega_0}}
}^2
\nn\\
\av{y_k\ooo(t)\ y_{k'}\ooo(t')}&=&\av{y_k\ooo{}\da(t)\ y_{k'}\ooo{}\da(t')}=0\ .
\end{eqnarray}

\subsection{
Conditions for the engineered Hamiltonian
}\label{Sec:necessaryConditions}

To reproduce, with our system, a dynamics similar to that of Ref.~\cite{zippilli2021}, we require the Hamiltonian~\rp{H} from the previous section to match the effective photon-mediated Hamiltonian~\rp{Heff}, that is
\begin{eqnarray}
\JJ=\JJ\al{eff}\ .
\end{eqnarray}
Specifically, by comparing Eqs.~\rp{JJ1}-\rp{vg0J} with \rp{JJ2}-\rp{WW2}, we find the necessary  conditions for the stabilization of a specific Gaussian steady state of the form of Eq.~\rp{Psi}, namely 
\begin{eqnarray}\label{vg0W}
\vg_0&=&\vg_0\al{\JJ}
\nn\\
\WW&=&\WW\al{\JJ}\ ,
\end{eqnarray}
that can be expressed as
\begin{eqnarray}\label{gJV}
\GG_{0,k}&=&-\ii\,J_0\ \ee^{-\ii\,\frac{\varphi_{1}-\varphi_0}{2}}\ \VV_{k,1}^*
\\
\WW\al{\delta}+\ovl\GG\da\ \DD\ \ovl\GG&=&\ii\ \VV\ \Phi\ \ovl\JJ\al{S}\ \Phi^*\ \VV\da\ .
\label{WJV}
\end{eqnarray}
These relations can be used to determine the values of the optomechanical couplings $\GG_{j,k}$~\rp{GGjk} necessary to achieve the expected dynamics (see Sec.~\ref{Sec:GG} below).

Before analyzing the conditions under which the optically induced mechanical noise is sufficiently small in Sec.~\ref{Sec:validity}, we first examine additional constraints on the achievable steady state imposed by the Hamiltonian that can be realized in an optomechanical system (see Sec.~\ref{Sec:constraints}). We then discuss how to determine the optomechanical couplings $\GG_{j,k}$  that solve Eqs.~\rp{gJV}-\rp{WJV} (see Secs.~\ref{Sec:GG} and \ref{Sec:controlpar}). 

\subsubsection{%
Additional constraints%
}\label{Sec:constraints}%

We note that the relations~\rp{gJV} and \rp{WJV} entail additional constraints on the matrix $\VV$ which determines also the state we aim to generate. 

First, we notice that all the entries of the vector $\vg_0$ [see Eqs.~\rp{vg0} and \rp{GGjk}] have the same phase determined by the phase of the driving field of the zero-th cavity mode. So, from Eq.~\rp{gJV} we find that all the elements on the first column of the matrix $\VV$ must have the same phase modulo $\pi$, i.e.
\begin{eqnarray}\label{argV}
\arg\pg{\VV_{k,1}}=\arg\pg{\VV_{k',1}}+n_{k,k'}\ \pi\ ,
\end{eqnarray}
for all $k,k'\in\pg{1,...N}$ and for $n_{k,k'}\in\mathds{Z}$.
Secondly, according to Eq.~\rp{WW}, the matrix $\WW$ is real and symmetric. 
This, with Eq.~\rp{WJV}, entails that
\begin{eqnarray}
0&=&\WW-\WW^*
=\WW-\WW^T
\nn\\&=&
\ii\pt{\VV\ \Phi\ \ovl\JJ\al{S}\ \Phi^*\ \VV\da
+\VV^*\ \Phi^*\ \ovl\JJ\al{S}\ \Phi\ \VV^T}\ ,
\end{eqnarray}
that is
\begin{eqnarray}\label{VVJ}
\ovl\JJ\al{S}\ \Phi\ \VV^T\ \VV\ \Phi+\Phi\ \VV^T\ \VV\ \Phi\ \ovl\JJ\al{S}=0\ .
\end{eqnarray}
Eqs.~\rp{argV} and \rp{VVJ} define two additional constraints on the matrix $\VV$~\rp{VV} and hence on the state that can be prepared in the stationary regime with this approach [see Eqs.~\rp{Uket0}, \rp{Up} and \rp{VV}].

\subsubsection{%
Determination of the optomechanical coupling strengths $\GG_{j,k}$%
}\label{Sec:GG}%

Let us now assume that we have identified the matrix $\VV$ corresponding to a given state and that fulfills the required conditions~\rp{argV} and \rp{VVJ} (for a given choice of $\JJ\al{S}$ and $\Phi$). The corresponding optomechanical couplings $\GG_{j,k}$ that are necessary to realize the dissipative dynamics of Ref.~\cite{zippilli2021}, can be evaluated using the relations in Eqs.~\rp{gJV} and \rp{WJV}.

The values of $\GG_{0,k}$ are directly given by Eq.~\rp{gJV}. 
The other values of $\GG_{j,k}$, with $j\neq 0$, are determined by Eq.~\rp{WJV}, namely they have to fulfill the relation $\ovl\GG\da\ \DD\ \ovl\GG =\WW\al{\JJ}-\WW\al{\delta}$, with the constraint that $\ovl\GG\da\ \DD\ \ovl\GG \in \mathds{R}^{N\times N}$. So, in order to determine the matrix $\ovl\GG$, 
we diagonalize $\WW\al{\JJ}-\WW\al{\delta}$ [see Eqs.~\rp{WWJ} and \rp{Wdelta}], which is real symmetric [as expected when $\rp{VVJ}$ is fulfilled], in terms of a real diagonal matrix $\Lambda$ and a real orthogonal $\TT$, such that
\begin{eqnarray}\label{WmWdelta}
\WW\al{\JJ}-\WW\al{\delta}
&=&\ii\, \VV\ \Phi\ \ovl\JJ\al{S}\ \Phi^*\ \VV\da-\WW\al{\delta}
\nn\\
&=&
\TT^T\ \Lambda\ \TT \ .
\end{eqnarray}
And thus, we find $\ovl\GG\da\ \DD\ \ovl\GG=\TT^T\ \Lambda\ \TT$. In particular, we set the number of optical modes $M$ to be equal to the number of non-zero elements of $\Lambda$ (i.e. non zero eigenvalues of $\WW\al{\JJ}-\WW\al{\delta}$), and we restrict $\Lambda$ and $\TT$ to the non-zero eigenvalues and corresponding eigenvectors only, by defining the corresponding reduced matrices 
$\ovl\Lambda\in\mathds{R}^{M\times M}$ and $\ovl\TT\in\mathds{R}^{M\times N}$, such that 
\begin{eqnarray}\label{TLT}
\ovl\GG\da\ \DD\ \ovl\GG&=&\ovl\TT^T\ \ovl\Lambda\ \ovl\TT\ .
\end{eqnarray}

Now, by properly selecting the diagonal matrix $\DD$ [through the system parameters, according to Eq.~\rp{DD}] we can make sure that $\DD^{-1}\ovl\Lambda>0$. Namely we can select the sign of each element of $\DD$, that is controlled by the signs of $\Delta_j$, to be equal to the sign of the corresponding element of $\ovl\Lambda$. In this way we can define the real diagonal matrix $\sqrt{\DD^{-1}\ \ovl\Lambda}$, and we find
\begin{eqnarray}\label{ovlGG}
\ovl\GG=\sqrt{\DD^{-1}\ \ovl\Lambda}\ \ \ovl\TT\ .
\end{eqnarray}
We describe a few specific examples in the following sections. 

We note that, in general, in order to make the matrix $\DD^{-1}\ovl\Lambda$ positive, one have to use both positive and negative values of $\Delta_j$. This may rise a concern regarding the stability of the optomechanical system, especially for the modes with $\Delta_j<0$ \cite{bowen2015}. We will show that when we achieve a good steady state preparation, the strong dissipation through the zero-th optical mode, is able to stabilize the whole system even when $\Delta_j<0$ for some $j$ (see Sec.~\ref{Sec:results}). This is due to the fact that while the zero-th optical mode is driven resonantly on the red-mechanical sideband (i.e. $\Delta_0=\omega_0$) so that the mechanical modes are strongly coupled to the squeezed dissipative bath, we select the detuning of the other modes to be very large in magnitude, such that (as discussed with in more details in Sec.~\ref{Sec:validity} below) the corresponding incoherent processes that would tend to pump energy into the system and make it unstable are very small and negligible.  

We also note that, if one can also control the mechanical frequencies, specifically the values of $\delta\omega_k$ in Eq.~\rp{omegak}, then it is possible to use only positive values of $\Delta_j$. In details, we assume that $\Delta_j>0$ for all $j$. Correspondingly the sign of the diagonal matrix $\DD$ is negative, see Eq.~\rp{DD}. Then we select the values of $\delta\omega_k$ [that define the matrix $\WW\al{\delta}$~\rp{Wdelta}], such that 
\begin{eqnarray}\label{Wdelta0}
\WW\al{\delta}&=&{\rm Diag}\pt{\WW\al{\JJ}}
+\id\ \delta_0
\end{eqnarray}
where ${\rm Diag}\pt{\WW\al{\JJ}}$ is the diagonal matrix with elements equal to the diagonal of $\WW\al{\JJ}$~\rp{WWJ}, and 
\begin{eqnarray}\label{delta0}
\delta_0={\rm max}\pg{{\rm eig}\pq{\WW\al{\JJ}-{\rm Diag}\pt{\WW\al{\JJ}}} }\ ,
\end{eqnarray}
where ${\rm eig}\pq{\MM}$ indicates the eigenvalues of a matrix $\MM$.

Correspondingly, the matrix $\WW\al{\JJ}-\WW\al{\delta}$, defined in Eq.~\rp{WmWdelta}, is seminegative definite, meaning that all the elements of $\ovl\Lambda$ in Eq.~\rp{TLT} are negative, so that $\sqrt{\DD^{-1}\ \ovl\Lambda}$ in Eq.~\rp{ovlGG} is real and $\ovl\GG$ is well defined. In this way, the concerns regarding the system stability are avoided. Nevertheless as shown in Sec.~\ref{Sec:results}, this gives no specific advantage, in the regime in which the steady state preparation is optimal.

\subsubsection{%
Control parameters%
}\label{Sec:controlpar}%

It is important to highlight that, in order to engineer the required array Hamiltonian [see Eqs.\rp{Heff}, \rp{JJ2} and \rp{vg0W}] which is necessary for the steady state control of the $N$ mechanical modes, we need to determine the values of the $N+1+N(N+1)/2$ parameters that define the interaction terms $\vg_0$ and $\WW$~\rp{vg0W}.
In fact, as discussed in Sec.~\ref{Sec:constraints}, the phases of all the elements of $\vg_0$ are equal, so that it consists of $N+1$ independent parameters, while $\WW$ is real and symmetric and therefore it contributes to $N(N+1)/2$ additional independent parameters. This requires controlling the same number of optomechanical parameters. For $M = N+1$ optical modes, the corresponding optomechanical couplings $\GG_{j,k}$ can in principle provide the required set of control parameters, provided that they can be tuned independently. In practice, this means that one must be able to adjust the bare optomechanical couplings $g_{jk}^0$ independently [see Eq.~\rp{GGjk}]. Achieving such independent control is not straightforward: depending on the physical implementation, the bare couplings may be interdependent and subject to various constraints. Nevertheless, in principle, one can always introduce additional cavity modes, and hence additional control parameters, so that, for sufficiently large $M$, it becomes possible to identify a set of couplings $\GG_{j,k}$ that satisfies Eq.~\rp{TLT},  even in the presence of implementation-specific constraints. This task may require a numerical approach, which we do not pursue here, and is beyond the scope of the present work. Instead, in the results presented below, we have assumed a minimal number of optical modes and applied Eq.~\rp{ovlGG}.

\subsection{%
Conditions for negligible optically mediated mechanical dissipation%
}\label{Sec:validity}%

Once the couplings $G_{j,k}$ have been fixed, we can use the full model to determine the steady state of the system. We expect the dynamics to approximate those of Ref.~\cite{zippilli2021} provided that both the intrinsic mechanical noise (at rate $\gamma_k n_{T,k}$) and the optically mediated mechanical noise [described by the matrix $\YY$~\rp{YY} and by the correlation functions~\rp{avyy}] remain much weaker than the dissipative dynamics induced by the squeezed reservoir. The latter is governed by the competition between the dissipation of the zeroth optical mode (at rate $\kappa_0$) and the coherent mechanical dynamics characterized by the effective coupling matrix $\WW$~\rp{WW}.

In particular, a necessary condition, for the effective approximation of the dynamics of Ref.~\cite{zippilli2021} with our system is that the optics-mediated phonon dissipation, described by Eq.~\rp{YY}, is much weaker than the coherent terms, described by Eq.~\rp{WW},
\begin{eqnarray}
\abs{\YY_{k,k'}}\ll\abs{\WW_{k,k'}}\ .
\end{eqnarray}
This can be achieved if, for example, $\gamma_k$ are sufficiently small and
\begin{eqnarray}\label{conditionToNeglectDissipation}
\abs{\kappa_j^2+\Delta_j^2-\omega_0^2}\gg 2\ \kappa_j\ \omega_0\ .
\end{eqnarray}
In the result section we fulfill this relation by working in the dispersive regime
\begin{eqnarray}\label{dispersive}
\abs{\Delta_j}\gg\kappa_j,\omega_0\ ,
\end{eqnarray}
such that
\begin{eqnarray}\label{WWYYa}
\WW_{k,k'}&\sim&\delta\omega_k\ \delta_{k,k'}-2\sum_{j=1}^M\ \frac{\GG_{j,k}^*\ \GG_{j,k'}}{\Delta_j}
\\
\YY_{k,k'}&\sim&\gamma_k\ \delta_{k,k'}+4\sum_{j=1}^M\ \frac{\GG_{j,k}^*\ \GG_{j,k'}\ \kappa_j\ \omega_0}{\Delta_j^3}\ .
\label{WWYYb}
\end{eqnarray}

\subsection{%
Preparation of a Gaussian cluster state
}\label{Sec:rectgraph}%

Let us now consider the preparation of a mechanical Gaussian cluster state (see Sec.~\ref{Sec:ClusterState0})
\begin{eqnarray}\label{Psicluster0}
\ke{\Psi_{\rm cluster}}=
\ee^{-\ii\sum_{k,k'=1}^N\ \KK_{k,k'}\ c_k\da\ c_{k'}}\ 
\prod_{k=1}^N
\ee^{\frac{z}{2}\pt{\ee^{\ii\varphi_k}\ b_k\da{}^2-\ee^{-\ii\varphi_k}\ b_k^2}}
\ke{0}
\end{eqnarray}
where the matrix $\KK$ is determined through the matrix $\VV=\ee^{-\ii\,\KK}$~\rp{VV} and is related to the adjacency matrix $\AAA$ (symmetric with nonzero elements equal to one and $\AAA_{k,k}=0$) and to the nullifiers~\rp{nulli} by the relation~\rp{VVPhiVV}
\begin{eqnarray}\label{VPhi2V}
\VV\ \Phi^2\ \VV^T
=
-\Theta\ \frac{\AAA-\ii\ \id}{\AAA+\ii\ \id}\ \Theta\ ,
\end{eqnarray}
where $\Phi$ and $\Theta$ are diagonal matrices with $\Phi_{k,k}=\ee^{\ii\,\frac{\varphi_k}{2}}$ and $\Theta_{k,k}=\ee^{-\ii\,\theta_k}$. Eqs.~\rp{Psicluster0} and ~\rp{VPhi2V} determine a generic cluster state~\cite{zippilli2020}. When we want to prepare this state with our optomechanical system, the matrix $\VV$ has to fulfill not only the relation~\rp{VPhi2V}, i.e. \rp{VVPhiVV}, but also the additional constraints expressed by Eqs.~\rp{argV} and \rp{VVJ}. It is possible to verify numerically that in the case of a rectangular graph (namely, when the nonzero elements of the adjacency matrix $\AAA$ correspond to the edges of a rectangular graph) with at least a side with an odd number of nodes (see App.~\ref{App:general-graphs} for considerations on more general graphs), the Eqs.~\rp{VVPhiVV}, \rp{argV} and \rp{VVJ} are fulfilled when the squeezing phases are zero 
\begin{eqnarray}\label{phiell}
\varphi_\ell&=&0\ , \hspace{1cm}{\rm for\ all}\ \ell\ ,
\end{eqnarray}
(such that $\Phi=\id$), the coupling parameters $J_k$ (with $k\neq0$) are equal to the same value
\begin{eqnarray}
J_k&=&J \hspace{1cm}{\rm for}\ k\in\pg{1,...N-1}\ ,
\end{eqnarray}
the values of the phases in the matrix $\Theta$ are equal to 
\begin{eqnarray}\label{thetak}
\theta_k&=&
\pt{k+1}\,\frac{\pi}{2}\ ,
\end{eqnarray}
and finally 
\begin{eqnarray}\label{VV0}
\VV_{k,k'}
&=&\ee^{-\ii\,k\,\pi}\ \pg{\sqrt{-\frac{\AAA-\ii\ \id}{\AAA+\ii\ \id}}}_{k,k'}
\ ,
\end{eqnarray}
see Eq.~\rp{VPhi2V}. Eqs.~\rp{Psicluster0}-\rp{VV0} define the target state that we aim to prepare in the result section. 

As discussed in Sec.~\ref{Sec:DissEngineering}, the matrices $\VV$ and $\Phi$ also determine the required array Hamiltonian~\rp{H} through the relations~\rp{JJ1}, \rp{WWJ} and \rp{vg0J}. In the case of the model of Ref.~\cite{zippilli2021}, having a Hamiltonian of this form guarantees that the state~\rp{Psicluster0} is the unique steady state (see Sec.~\ref{Sec:DissEngineering}).

Furthermore, the effective phonon Hamiltonian~\rp{Heff} is equal to the expected array Hamiltonian~\rp{H} when the optomechanical coupling strengths $\GG_{j,k}$~\rp{GGjk} fulfill Eqs.~\rp{gJV} and \rp{WJV}. This, in turn, means that when these relations hold, and if additional uncontrolled noise is negligible, the system steady state closely approximate the target state~\rp{Psicluster0}.
In particular, the specific values of $\GG_{j,k}$ for the steady state preparation of a rectangular mechanical cluster state can be determined using the parameters~\rp{phiell}-\rp{VV0} in Eqs.~\rp{gJV} and \rp{WJV}, following the procedure outlined in Sec.~\ref{Sec:GG}.

\begin{figure*}[th!]
\centering
\includegraphics{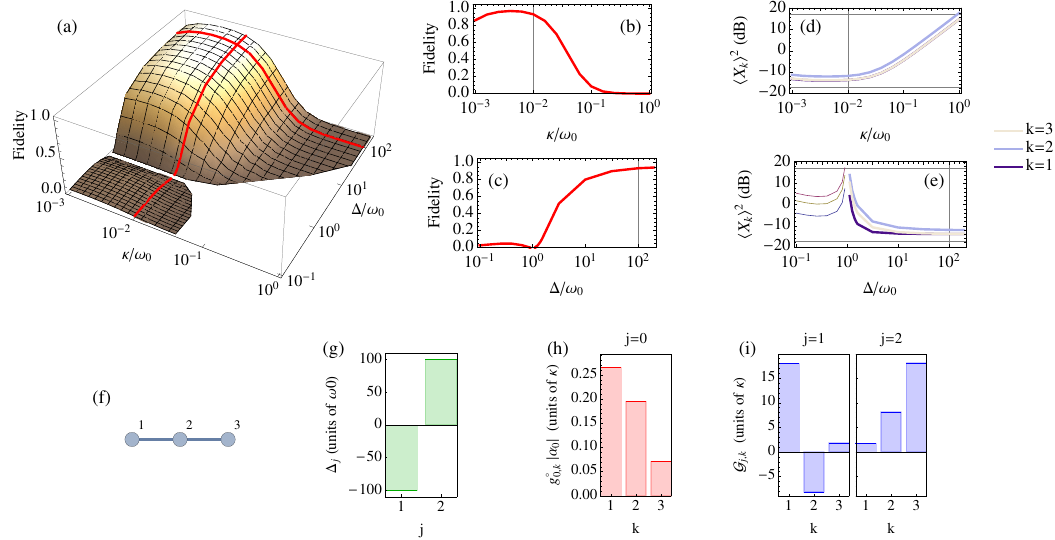}
\caption{
Fidelity (a)-(c) [Eq.~\rp{F}] and variance of the nullifiers (d)-(e) [Eq.~\rp{avX}], as a function of the optical decay rate $\kappa=\kappa_j$ for all $j$ (the same for all the optical modes) and of the optical detuning $\Delta=\abs{\Delta_j}$ for all $j$ [the same amplitude for all the optical modes but different signs as reported in (g)], for the preparation of a Gaussian cluster state corresponding to a graph with $N=3$ modes on a line (f).
This is the graph corresponding to the adjacency matrix $\AAA$ introduced in Sec.~\ref{Sec:ClusterState0} and discussed in Sec.~\ref{Sec:rectgraph}: the edges of the graph correspond to the nonzero elements of $\AAA$.
The results in (b)-(e) are evaluated for the values of $\kappa$ and $\Delta$ corresponding to the red lines in (a). These results are evaluated for $J_0=3.4\times 10^{-3}\omega_0$ and $J_\ell=J=0.6\times 10^{-3}\omega_0$ for $\ell\in\pg{1,...N}$, [see Eqs.~\rp{JJS} and \rp{ovlJJS}], which we have found by maximizing the fidelity, as a function of $J_0$ and $J$, for the specific values of $\kappa$ and $\Delta$ indicated by the red lines in (a) and by the vertical lines in (b)-(e); (g) shows the values of $\Delta_j$ used for this maximization: the amplitude for all the modes is equal to the value indicated by the vertical line in (c), but the sign can be different. For these specific values of $\kappa$ and $\Delta$, the values of the optomechanical couplings $\GG_{j,k}$ are reported in (h) and (i). The values in (h) determine $\GG_{0,k}$ [see Eq.~\rp{GGjk}] according to $\GG_{0,k}=g_{0,k}\oo\,\abs{\alpha_0}\ \ee^{\ii\,{\rm arg}\pq{\alpha_0}}$, and here ${\rm arg}\pq{\alpha_0}=-\pi/2$. The other parameters are $\gamma_k=10^{-8}\omega_0$ for all $k$, T=0.01K, $\omega_0=1$GHz, $\delta\omega_k=0$ for all $k$ (i.e. resonant mechanical modes), and squeezing parameter of the reservoir $r=2$ [see Eq.~\rp{corrSqueezBath}]. Regions in (a) where the surface plot is missing indicate parameter regimes where the system is unstable.
}
\label{fig1}
\end{figure*}

\begin{figure*}[t!]
\centering
\includegraphics{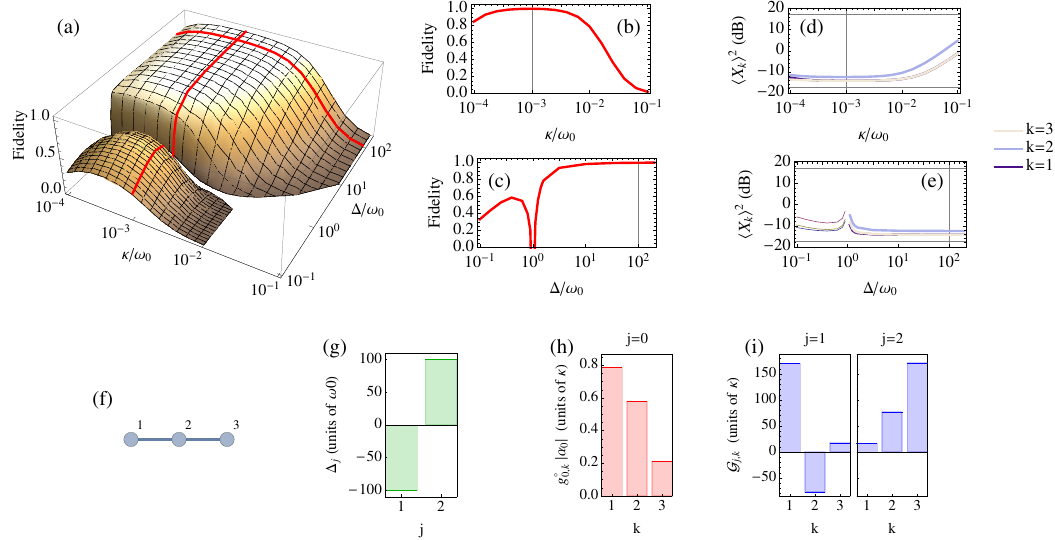}
\caption{
As in Fig.~\ref{fig1} but for  $J_0=10^{-3}\omega0$, and $J=0.5\times10^{-3}\omega_0$, which have been found by maximizing the fidelity, as a function of $J_0$ and $J$, for a smaller value of $\kappa$, identified by the vertical line in (b).
The values of the other parameters are as in Fig.~\ref{fig1}.
}
\label{fig2}
\end{figure*}

\begin{figure*}[t!]
\centering
\includegraphics{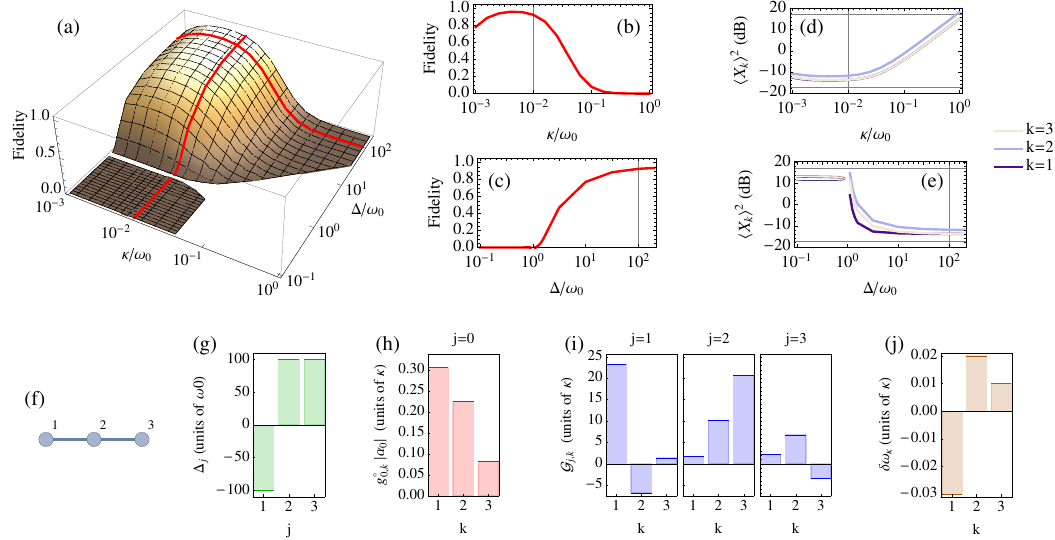}
\caption{As in Fig.~\ref{fig1}, but for non-resonant mechanical resonators. The values of $\delta\omega_k$ are reported in (j), they are chosen so that the average mechanical frequency $\sum_k\omega_k/N=\omega_0$. Here $J_0=3.9\times 10^{-3}\omega_0$ and $J=0.7\times 10^{-3}\omega_0$.
The values of the other parameters are as in Fig.~\ref{fig1}.}
\label{fig3}
\end{figure*}

\begin{figure*}[t!]
\centering
\includegraphics{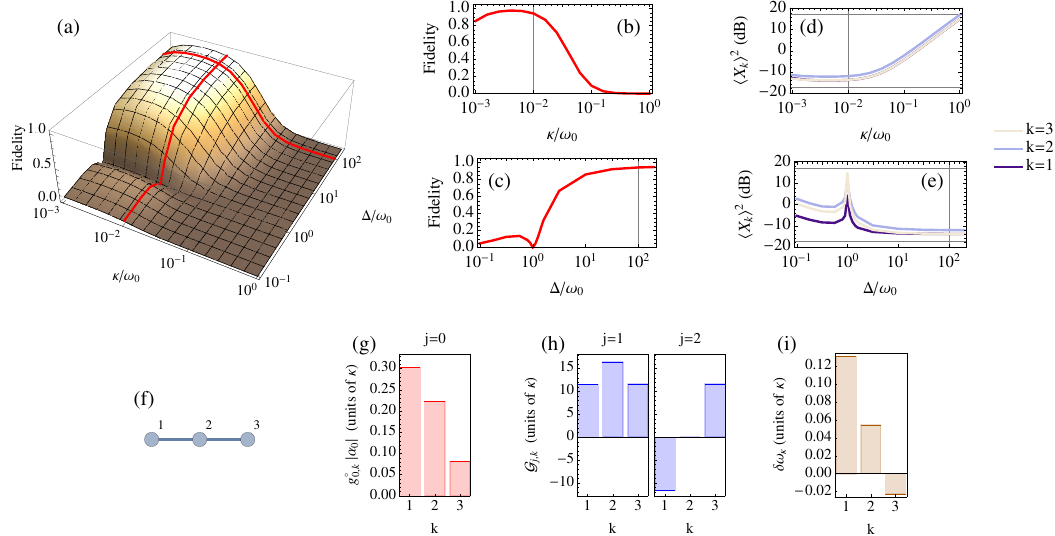}
\caption{As in Fig.~\ref{fig1}, but evaluated by selecting the specific values of $\delta\omega_k$ (i), for which we can set all the optical detuning equal positive ($\Delta=\Delta_j$ for all $j$), as discussed in Sec.~\ref{Sec:GG}. Here $J_0=3.8\times 10^{-3}\omega_0$ and $J=0.7\times 10^{-3}\omega_0$. In this case, the system is stable over the whole plotted region.
The values of the other parameters are as in Fig.~\ref{fig1}.
}
\label{fig4}
\end{figure*}

\begin{figure*}[t!]
\centering
\includegraphics{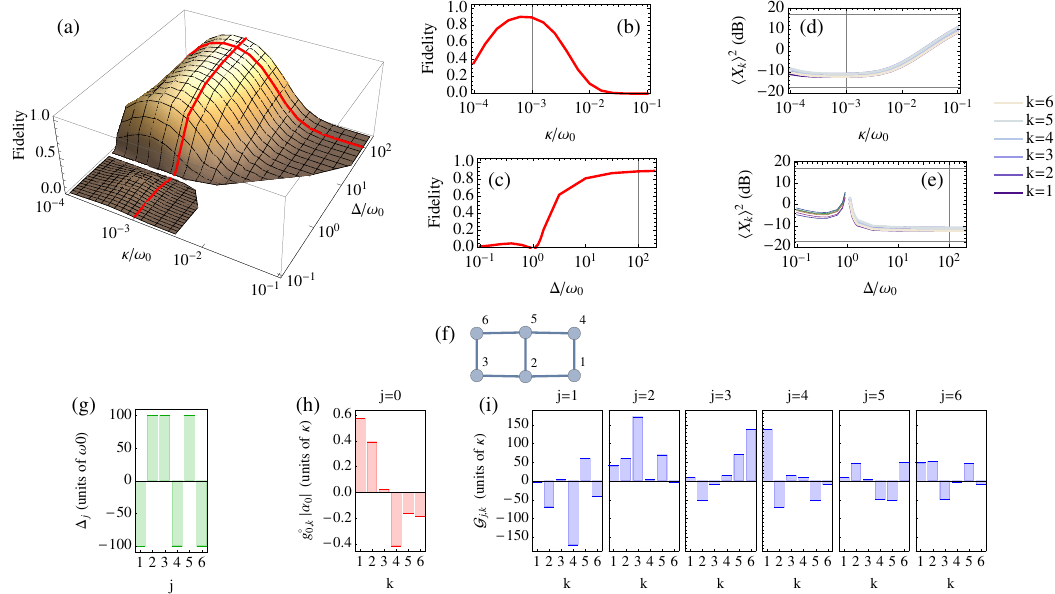}
\caption{
As in Fig.~\ref{fig2} but for a rectangular graph with $N=2\times 3=6$ mechanical modes (f). Here $J_0=8.5\times 10^{-4}\omega_0$ and $J=4.4\times 10^{-4}\omega_0$.
The values of the other parameters are as in Fig.~\ref{fig2}.
}
\label{fig5}
\end{figure*}

\begin{figure*}[t!]
\centering
\includegraphics{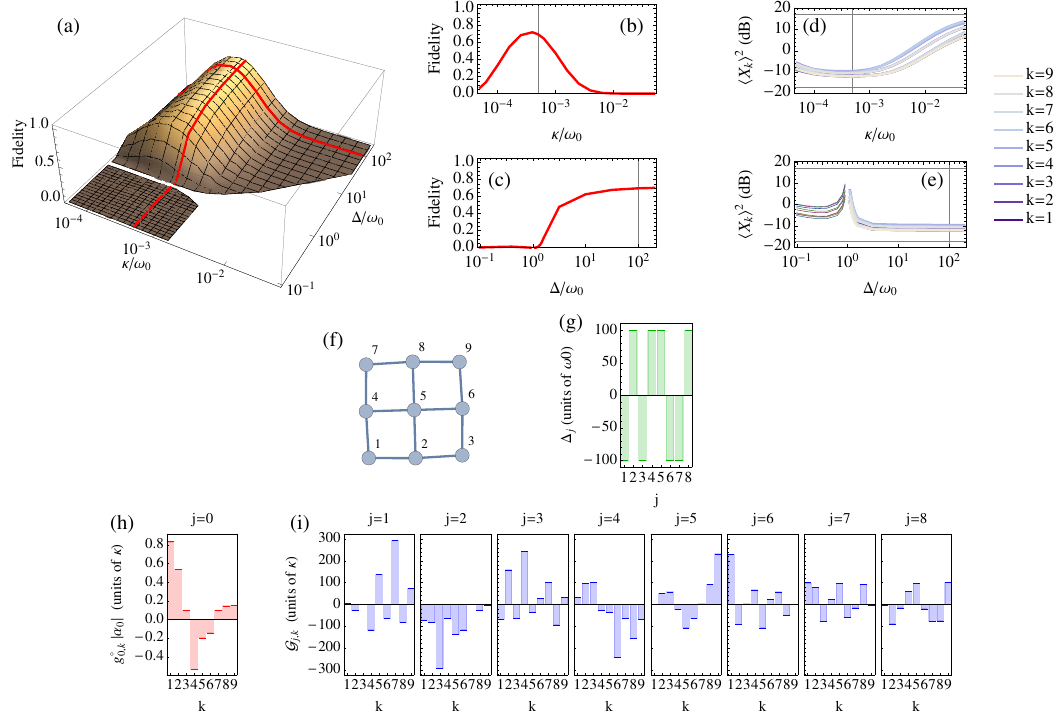}
\caption{
As in Fig.~\ref{fig2} but for a rectangular graph with $N=3\times 3=9$ mechanical modes (f). Here $J_0=6\times 10^{-4}\omega_0$ and $J=3.5\times 10^{-4}\omega_0$.
The values of the other parameters are as in Fig.~\ref{fig2}.
}
\label{fig6}
\end{figure*}

\begin{figure*}[t!]
\centering
\includegraphics{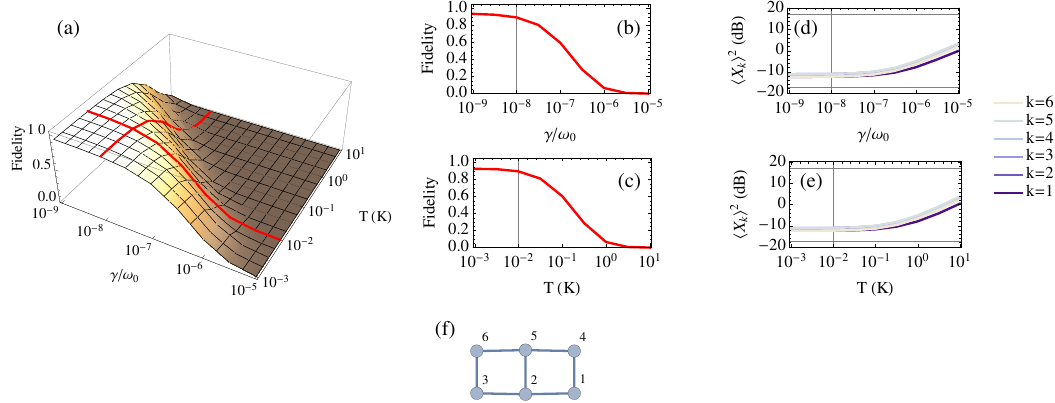}
\caption{
Fidelity (a)-(c) and variance of the nullifiers (d)-(e), as a function of the mechanical decay rate $\gamma=\gamma_k$, for all $k$, and of the temperature  of the mechanical resonators $T$, for the preparation of a Gaussian cluster state corresponding to a graph with $N=6$ modes on $2\times 3$ rectangular graph (f). 
The other parameters are as in Fig.~\ref{fig5}.
}
\label{fig7}
\end{figure*}

\begin{figure}[t!]
\centering
\includegraphics[width=0.48\textwidth]{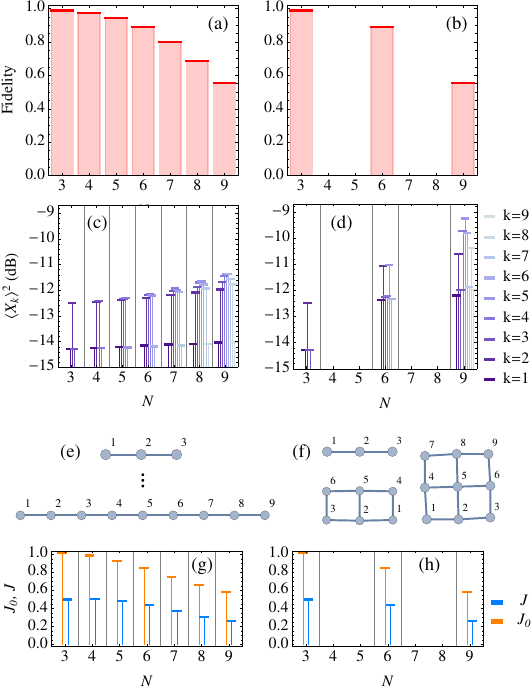}
\caption{%
Fidelity (a)-(b) and variance of the nullifiers (c)-(d) for the preparation of Gaussian cluster states corresponding to linear graphs (left column) with $N=\pg{1,...9}$ (e) and to square graphs (right column) with $N=\pg{3,6,9}$ (the results with $N=3$ are the same in both columns).
All results are evaluated for $\kappa=10^{-3}\omega_0$ as in Figs.~\ref{fig2} and \ref{fig5} [i.e., the results with $N=3$ and $N=6$ in the right column are equal to the result at the crossing point between the red lines in Figs.~\ref{fig2} (a) and \ref{fig5} (a)]. 
The optomechanical couplings are determined following the procedure detailed in Sec.~\ref{Sec:GG}, with the values of $J_0$ and $J$ reported in panels (g) and (h). The other parameters are as in Figs.~\ref{fig1}.
}
\label{fig8}
\end{figure}

\begin{figure*}[t!]
\centering
\includegraphics[width=0.8\textwidth]{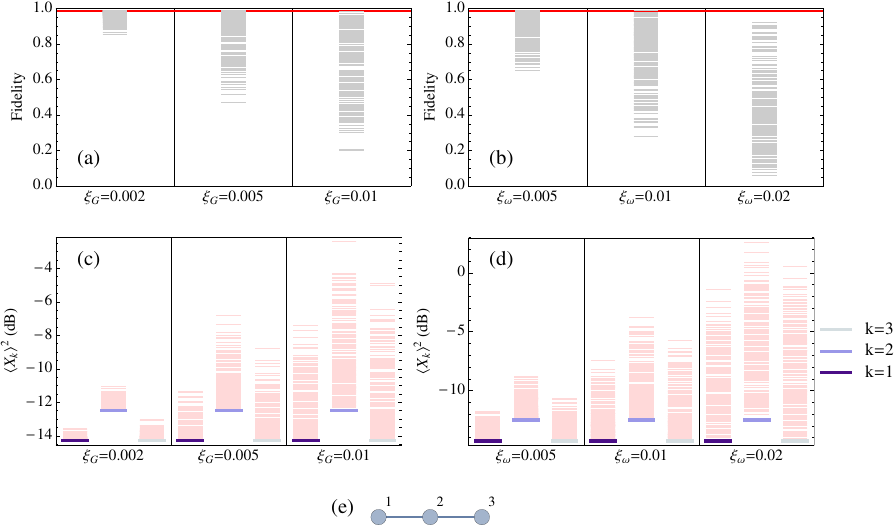}
\caption{%
Fidelity (a)-(b) and variance of the nullifiers (c)-(d) for the preparation of Gaussian cluster states corresponding to linear graphs with $N=3$ (e). The thick red lines in (a)-(b) and the thick blue lines in (c)-(d) are equal to the corresponding result in Fig.~\ref{fig8}. 
The thin gray lines in (a) and the thin light-red lines in (c) correspond to 200 random realizations evaluated using the same parameters as the thick lines and setting the optomechanical couplings to the values $\GG_{j,k}\pt{1+\xi_{j,k}\al{G}}$, where $\xi_{j,k}\al{G}$ are random variables uniformly distributed in the range $\xi_{j,k}\al{G}\in\pq{-\xi_G,\xi_G}$. 
The thin gray lines in (b) and light-red lines in (d) correspond to 200 random realizations evaluated using the same parameters as the thick lines and setting the mechanical frequencies to the values $\omega_k+\kappa\,\xi_{j,k}\al{\omega}$, where $\xi_{j,k}\al{\omega}$ are random variables uniformly distributed in the range $\xi_{j,k}\al{\omega}\in\pq{-\xi_\omega,\xi_\omega}$. The specific values of $\xi_G$ and $\xi_\omega$ are indicated in the lower parts of each plot.  
}
\label{fig9}
\end{figure*}
  
\begin{figure*}[t!]
\centering
\includegraphics[width=0.8\textwidth]{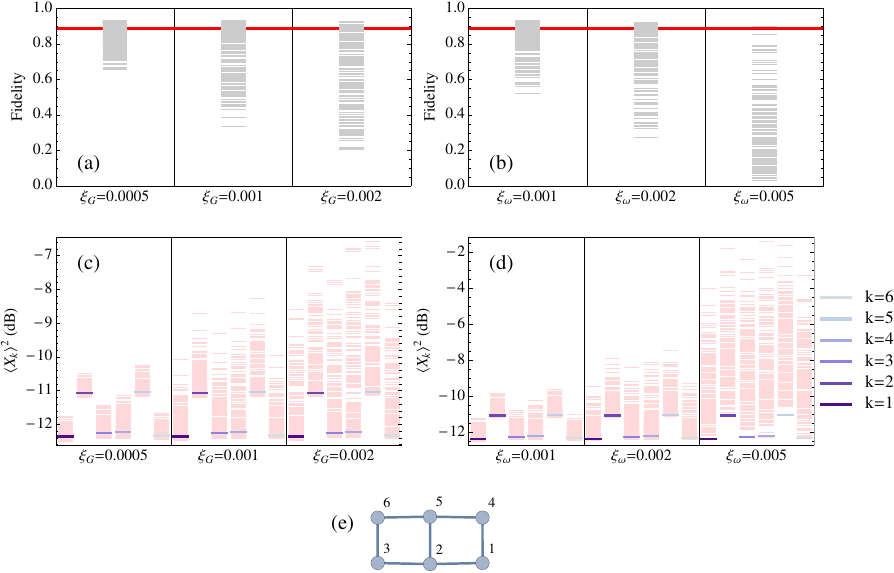}
\caption{%
As in Fig.~\ref{fig9} but evaluated for the the parameters of Fig.~\ref{fig8} corresponding to the $2\times 3$ rectangular graph.
}
\label{fig10}
\end{figure*}

\section{Results}
\label{Sec:results}

We now verify the validity of our previous analysis, showing that the steady state of the optomechanical system governed by the quantum Langevin equations~\rp{QLE00} indeed corresponds to a cluster state when the coupling parameters are chosen according to the procedure outlined in Sec.~\ref{Sec:approx}. 

Before analyzing the numerical results, we summarize the conditions under which we expect to observe the dynamics of Ref.~\cite{zippilli2021}, which enables the dissipative stabilization of rectangular Gaussian cluster states, as discussed in the previous section.
\begin{itemize}
\item[(a)]
The mechanical resonators are assumed to be nearly resonant, $\delta\omega_k\ll\kappa_j$, so that the dominant mechanical processes are those that conserve the number of mechanical excitations; processes that do not conserve the number of mechanical excitations are off-resonant and negligible;

\item[(b)]
We assume fast dynamics of the optical modes with $j\neq~0$, such that
\begin{eqnarray}
\abs{\kappa_j+\ii\,\Delta_j}\gg\GG_{j,k}\ , \ \ \ \ {\rm for}\ j\neq 0\ ;
\nn
\end{eqnarray}
This allows for the adiabatic elimination of these modes, resulting in effective optically mediated phonon-phonon interactions and optically induced dissipation;

\item[(c)]
Weak optically induced dissipation requires operation in the dispersive regime, according to Eq.~\rp{dispersive}:
\begin{eqnarray}
\abs{\Delta_j}\gg\kappa_j,\ \omega_0\ , \ \ \ \ {\rm for}\ j\neq 0\ ;
\nn
\end{eqnarray}

\item[(d)]
The zeroth optical mode is pumped on the red mechanical sideband, $\Delta_0= \omega_0$;
This provides an efficient indirect coupling between the zeroth mode and the mechanical resonators;

\item[(e)] 
To obtain an engineered Hamiltonian equal to the one required by the theorem of Ref.~\cite{zippilli2021}, namely the Hamiltonian in Eq.~\rp{H(V)}, the optomechanical couplings must satisfy Eqs.~\rp{gJV} and \rp{WJV}, namely
\begin{eqnarray}
\GG_{0,k}
&=&
-\ii\,J_0\ \ee^{-\ii\,\frac{\varphi_{1}-\varphi_0}{2}}\ \VV_{k,1}^*
\ ,
\nn
\\
\WW\al{\delta}+\ovl\GG\da\ \DD\ \ovl\GG
&=&
\ii\ \VV\ \ovl\JJ\al{S}\ \VV\da\ ,
\nn
\end{eqnarray}
where $\VV$ depends on the specific state we aim to stabilize (see below) (and where we set the phases in $\Phi$ to zero, which is always allowed, see App.~\ref{App:general-graphs}); The solutions of these equations can be found following the procedure discussed in Secs.~\rp{Sec:constraints}, \rp{Sec:GG} and \rp{Sec:controlpar};

\item[(f)] 
In addition, the mechanical thermal noise must be small, i.e., much smaller than both the coherent phonon–phonon couplings [see Eq.~\rp{WWYYa}] and the dissipation rate in the squeezed bath $\kappa_0$:
\begin{eqnarray}
\gamma_k\ n_{T,k} \ll 
\abs{
\frac{\GG_{j,k}\GG_{j,k'}}{\Delta_j}
},\ 
\kappa_0\ ;
\end{eqnarray}

\item[(g)] 
Finally, we focus on rectangular Gaussian cluster states, which require a specific form of the required Hamiltonian~\rp{H(V)}, or specifically of the matrix $\VV$. 
At the same time, our physical system imposes constraints (see Sec.~\ref{Sec:constraints}) on the state that we want to stabilize. In the case of rectangular cluster states, these constraints correspond to fixing specific phases $\theta_k$ of the nullifiers [see Eq.~\rp{nulli}].  In particular, for a rectangular Gaussian cluster state with adjacency matrix $\AAA$, according to Eqs.~\rp{thetak} and \rp{VV0},
\begin{eqnarray}
\VV_{k,k'}
&=&
\ee^{-\ii\,k\,\pi}\ \pg{\sqrt{-\frac{\AAA-\ii\ \id}{\AAA+\ii\ \id}}}_{k,k'}
\ ,
\nn\\
\theta_k
&=&
\pt{k+1}\,\frac{\pi}{2}\ .
\nn
\end{eqnarray}

\end{itemize}
Specifically, we study the preparation of rectangular cluster states of various dimensions. For each case, we determine the required optomechanical couplings solving Eqs.~\rp{gJV} and \rp{ovlGG} with the parameters defined by Eqs.~\rp{phiell}–\rp{VV0}. We then compute the steady-state correlation matrix of the full model~\rp{QLE00} using standard techniques (see App.~\ref{App:Standard method}). To assess the performance of our scheme, we evaluate two key figures of merit: (i) the fidelity between the steady state of the full dynamics and the target cluster state, and (ii) the variances of the nullifiers associated with the desired cluster state, computed with respect to the steady state of the full model (see App.~\ref{App:fidelity-nullifiers}). 
The fidelity tells how close two states are~\cite{spedalieri2013}, while the variance of the nullifiers quantify the strength of the entanglement of the cluster state (the smaller the variance the stronger the entanglement between the various modes) and is a valuable metric to establish whether the state is usable for fault-tolerance measurement based quantum computation~\cite{menicucci2014,fukui2018,walshe2019}. 
We study systems containing between $6$ modes (the smallest case with N=3, Figs.~\ref{fig1}, \ref{fig2}, and \ref{fig4}) and $18$ modes (the largest case with N=9, Fig.~\ref{fig6}), see Fig.~\ref{fig8}. For each configuration, we compute the steady-state covariance matrix of the full model, then extract the reduced covariance matrix of the mechanical modes alone, from which the above quantities are evaluated. The result presented hereafter are evaluated by setting the detuning of the zero-th mode to $\Delta_0=\omega_0$, indicating that this mode is driven on the red mechanical sideband, and providing a relatively efficient indirect coupling between the squeezed reservoir and the mechanical resonators~\cite{asjad2016a}. Moreover, the mechanical frequencies are $\omega_k\sim 1GHz$ consistent with recent multi-mode opto- and electromechanical experiments~\cite{
mercade2021,
fiaschi2021,
kharel2022,
madiot2023a,
andersson2022,
qiao2023,
vonlupke2024%
}. In Figs.~\ref{fig1}-\ref{fig6} and \ref{fig8}-\ref{fig10} the quality factor is $\omega_k/\gamma_k\sim10^8$ and the temperature $T=10$\,mK. These are relatively demanding values which indicate that the dynamics we are describing is quite sensitive to thermal noise. 

Figs.~\ref{fig1}-\ref{fig6}, are evaluated as follows. First we selected specific values of the cavity decay rate $\kappa_j$ (relatively small, so that the corresponding noise induced on the mechanical dynamics is weak) and of the optical detuning $|\Delta_j|$ (relatively large, so that the relation~\rp{conditionToNeglectDissipation} is fulfilled). For simplicity both values are assumed equal for all the optical modes and we use the symbols $\kappa\equiv\kappa_j$ and $\Delta\equiv\abs{\Delta_j}$ for $j\in\pg{1,...M}$. These are the values identified by the red lines in the surface plots and by the vertical lines in the other plots. 
Then, the values of the coupling constant $J_\ell$, for $\ell\in\pg{0,...N}$, that constitute the auxiliary chain Hamiltonian~\rp{HS}, and determine the expected ideal  Hamiltonian (see Sec.~\ref{Sec:DissEngineering}),  are chosen  as the values that maximize the fidelity as a function of $J_\ell$, and at fixed $\kappa$ and $\Delta$. In particular here we have selected all the couplings with $\ell>0$ equal to the same value $J\equiv J_\ell$ for $\ell\in\pg{1,...N}$. In this way, the maximization is performed over the two parameters $J_0$ and $J$. Finally, the shown plots are found for these specific values of $J_0$ and $J$, by varying the values of $\kappa$ and $\Delta$ themselves. Since the values of the optomechanical couplings $\GG_{jk}$ depend on both the cavity parameters ($\kappa$, $\Delta$) and on the parameters that determine the expected Hamiltonian ($J_0$ and $J$), 
the values of $\GG_{jk}$ are determined and actualized at every different value of $\kappa$, $\Delta$, $J_0$ and $J$ in all our calculations (i.e. both in the maximization process and in the evaluation of the plots).

In general, we observe that optimal preparation is achieved at large detuning $\Delta$ and relatively small but finite $\kappa$ (in the resolved sideband limit). Comparing Figs.~\ref{fig1} and \ref{fig2}, we note that it is possible to achieve good preparation for smaller and smaller values of $\kappa$ provided also the values of $J_j$ are properly adjusted. In fact, in this case the preparation results even more efficient. This is due to the fact that overall the relative strength between the optically induced coherent couplings [described by Eqs.~\rp{WW} and \rp{DD}] and the optically induced dissipation [described by Eq.~\rp{YY}], in fact increases with decreasing $\kappa$ [see Eqs.~\rp{WWYYa}-\rp{WWYYb}]. 
In any case this improvement is feasible only if $\kappa$ is not too small. At very small $\kappa$ the system dynamics is inevitably too slow to overcame natural mechanical thermal decoherence.

Figs.~\ref{fig1}-\ref{fig4} show the results for the simplest case with $N=3$
mechanical resonators. In Fig.~\ref{fig1} and \ref{fig2}, we assumed the mechanical oscillator to be resonant with frequency equal to $\omega_0$, i.e. $\delta\omega_k=0$, see Eq.~\rp{omegak}. In this symmetric case the   matrix $\WW\al{\JJ}-\WW\al{\delta}$ [see Eq.~\rp{WmWdelta}] has a zero eigenvalue and as a consequence, two additional optical modes (excluding the zero-th mode) are sufficient to engineer the effective Hamiltonian for the three mechanical modes. If the mechanical resonators have different frequencies, in general, the matrix $\WW\al{\JJ}-\WW\al{\delta}$ has no zero modes and, one should use an additional optical mode to control the mechanical dynamics. An example of this situation is reported in Fig.~\ref{fig3}, where the frequencies are chosen so that the average mechanical frequency is equal to the frequency used in Figs.~\ref{fig1} and \ref{fig2}. In any case, we observe that, as far as the frequency differences between the mechanical modes is much smaller than the cavity decay rate $\kappa$, the state preparation is still good.

In the surface plots of Figs.~\ref{fig1}-\ref{fig3}, \ref{fig5}, and \ref{fig6}, the white areas correspond to parameter values for which the system becomes unstable. Specifically, in these regions, at least one of the eigenvalues of the drift matrix [i.e. the matrix of coefficients of the system of equations~\rp{QLE00}] has a positive real part. For these parameters, the linearized Eqs.~\rp{QLE00} are not valid and cannot be used to determine the long time behavior of the system (see App.~\ref{App:Standard method}). These instability regions can appear when the optical detunings are negative, such that light-induced heating processes dominate over cooling ones~\cite{bowen2015}. As explained in Sec.~\ref{Sec:GG}, negative detunings are needed to match the signs of the eigenvalues of the matrix $\WW\al{\JJ}-\WW\al{\delta}$~\rp{WmWdelta}. However, if we can control also the mechanical frequencies we can select specific mechanical frequencies to make the $\WW\al{\JJ}-\WW\al{\delta}$ seminegative definite [see Eqs.~\rp{Wdelta0} and \rp{delta0}] and to both reduce the number of the optical modes by having zero eigenvalues and avoid any instability by using only positive values of $\Delta_j$. This is described by Fig.~\ref{fig4}. In any case, this gives no specific advantage for parameters corresponding to the largest fidelity, which remains at levels analogous to the other cases. 

When we increase the size of the cluster state the preparation becomes more difficult (see Figs.~\ref{fig5}-\ref{fig7}). Larger fidelities are achievable only if thermal effects are reduced by either reducing the mechanical natural decay rate $\gamma$ or the temperatures, see Fig.~\ref{fig7}. In any case, we observe, in Figs.~\ref{fig5} and \ref{fig6}, that even if the fidelities don't reach very large values, the variance of the nullifiers (see App.\ref{App:avX}) is still at the same level of the other plots. This indicates that even if the state is not exactly equal to the expected cluster state we aim to generate, its entanglement properties (that are relevant for task such as measurement based quantum computation~\cite{yazdi2024}) are less sensitive to the system size.

The scaling with the system size is analyzed in more details in Fig.~\ref{fig8}. Here, we study the behavior of the fidelity and the variance of the nullifiers for linear and rectangular graphs. 
In general, the state preparation becomes less efficient as the array size increases. Moreover, the fidelity exhibits similar scaling for both geometries, while the variance of the nullifiers is more strongly affected in the square geometry, which has more connections and a more complex entanglement structure. 

Finally, in Figs.~\ref{fig9} and \ref{fig10} we analyze the stability of the result under small random variation of some system parameters. In particular, we vary the optomechanical couplings and the mechanical frequencies, and we notice that larger arrays are more sensitive to these random imperfections. Interestingly, Fig.~\ref{fig10} shows that it is even possible to tweak slightly the couplings and the frequencies to achieve better preparation (i.e. large fidelity and smaller variance of the nullifiers) than that achieved by following exactly our protocol. 

\section{Conclusions}
\label{conclusions}

We have demonstrated how the quantum steady state of a multimode optomechanical system can be controlled using a single squeezed reservoir. The system consists of multiple mechanical and optical modes, with the squeezed reservoir coupled to a single optical mode. By tuning the optomechanical interaction strengths, one can engineer an effective photon-mediated phonon–phonon interaction Hamiltonian with the properties required for the quantum state preparation protocol of Ref.~\cite{zippilli2021}. This protocol enables the dissipative stabilization of complex quantum steady states of the mechanical modes, including Gaussian cluster states~\cite{zhang2006,zippilli2020}. 

We presented results for a squeezed reservoir with large squeezing bandwidth. In the resolved sideband limit ($\kappa\ll\omega_k$), which we studied in this work, large squeezing bandwidth means that the bandwidth has to be much larger than the frequency range spanned by the normal modes of the effective mechanical array. In this way all the normal modes experience roughly the same degree of squeezing and the infinite bandwidth model used here is valid (see similar considerations in a related setting in Ref.~\cite{zippilli2014}). For a smaller bandwidth, the transfer of quantum correlations from the reservoir to the system is inevitably degraded, and the preparation of the target state is expected to be less efficient~\cite{zippilli2014,asjad2016a}.  

In particular, we have shown that high-fidelity rectangular Gaussian cluster states of the mechanical modes can be prepared using mechanical oscillators with GHz frequencies and quality factors of the order of $10^8$. Such parameters are either already within reach of current experimental technology or are expected to become accessible in the future~\cite{maccabe2020,ren2022a}. Mechanical resonators in the GHz range with sufficiently high quality factors have already been demonstrated~\cite{maccabe2020}. To observe the dynamics discussed here, similar performance levels will need to be achieved in devices that also support multi-mode operation~\cite{ren2022a}.
In fact, reported experiments with multimode optomechanical system are still far from the parameter regime needed to demonstrate the dynamics discussed in the present work. Nevertheless, recent progress in this direction has been reported. For example, Ref.~\cite{chegnizadeh2024} demonstrates quantum behavior in the collective dynamics of six mechanical oscillators, but at MHz frequencies and with quality factors around $10^7$.

\acknowledgments{We acknowledge the support of PNRR MUR project PE0000023-NQSTI (Italy).}

\appendix

\section{Evaluation of the effective model}
\label{App:EffectiveModel}

In order to estimate the values of the coupling constant necessary to realize the dynamics of Ref.~\cite{zippilli2021} we determine the effective phonon-phonon interaction strengths  by adiabatically eliminating the optical fields with $j\neq 0$. Specifically, we consider Eq.~\rp{QLE00} and we assume that condition~\rp{cond_AdEl} is fulfilled, meaning that 
the optical dynamics is much faster than that of the slowly varying mechanical operators
\begin{eqnarray}
b_k\ooo(t)=\ee^{\ii\ \omega_0\ t}\ b_k(t)\ .
\end{eqnarray}
Accordingly we consider the steady value of the optical operators, with $j\neq 0$,
\begin{eqnarray}\label{ajsteady}
a_j(t)&=&a_j(t)\Bigl|_{\GG=0}+F\pq{\wt b_k(\omega),\wt b_k\da(\omega)}
\end{eqnarray}
where, with the symbol $\wt{\ }$ we indicate quantities in Fourier space
$
\wt x(\omega)\equiv\frac{1}{\sqrt{2\,\pi}}\int\, \dd\,t\ \ee^{\ii\,\omega\,t}\ x(t)
$, 
we introduced the steady state optical operators in the absence of the mechanical modes
\begin{eqnarray}
a_j(t)\Bigl|_{\GG=0}&=&
\sqrt{\frac{\kappa_j}{\pi}}
\int_{-\infty}^\infty\, \dd\,\omega\ \ee^{-\ii\,\omega\,t}\ 
\wt\chi_j(\omega)\ \wt a_j\al{in}(\omega)
\ ,
\end{eqnarray}
with
$\wt\chi_j(\omega)$ the cavity susceptibility
\begin{eqnarray}
\wt\chi_j(\omega)&=&\frac{1}{\kappa_j+\ii\pt{\Delta_j-\omega}}\ ,
\end{eqnarray}
and where the interaction with the mechanical modes is described by the term
\begin{eqnarray}
F\pq{\wt b_k(\omega),\wt b_k\da(\omega)}&=&
-\frac{\ii}{\sqrt{2\,\pi}}\int_{-\infty}^\infty\dd\omega\ \ee^{-\ii\,\omega\,t}\ \wt{\XX}_j(\omega)
\nn\\&&\hspace{0cm}\times 
\sum_{k=1}^N\GG_{j,k}\ \pq{\wt b_k(\omega)+\wt b_k\da(\omega)}\ .
\end{eqnarray}
Then we approximate Eq.~\rp{ajsteady} and the corresponding result for the creation operators, as described in the following, and substitute them in the equation for $b_k\ooo(t)$ by keeping only  the resonant terms, and eventually we find
Eq.~\rp{QLE01} of the main text.

Specifically, we approximate $F\pq{\wt b_k(\omega),\wt b_k\da(\omega)}$ assuming that the slowly varying mechanical operators $b_k\ooo(t)$ are essentiually constant over the cavity dynamics, such that
\begin{eqnarray}
F\pq{\wt b_k(\omega),\wt b_k\da(\omega)}&\simeq&
-\ii\int_{-\infty}^\infty\dd\omega\ \ee^{-\ii\,\omega\,t}\ \wt{\XX}_j(\omega)
\\&&\hspace{-2.5cm}\times 
\sum_{k=1}^N\GG_{j,k}\ \pq{
\delta(\omega-\omega_0)\ b_k\ooo(t)
+
\delta(\omega+\omega_0)\ b_k\ooo{}\da(t)
}
\nn\\&&\hspace{-2.5cm}=
-\ii\sum_{k=1}^N\GG_{j,k}\pq{
\wt\chi_j(\omega_0)\ \ee^{-\ii\,\omega_0\,t}\ b_k\ooo(t)
+
\wt\chi_j(-\omega_0)\ \ee^{\ii\,\omega_0\,t}\ b_k\ooo{}\da(t)
}\, .
\nn
\end{eqnarray}
Thereby the equation for $b_k\ooo(t)$ can be approximated, keeping only  the resonant terms, as
\begin{eqnarray}
\dot b_k\ooo&=&-\sum_{k'=1}^N\pt{\gamma_k\ \delta_{k,k'}+\ii\,\delta\omega_k\ \delta_{k.k'}+\ii\ \FF_{k,k'}}\,b_{k'}
-\ii\ \GG_{0,k}^*\,a_0\ooo
\nn\\&&
+ \sqrt{2\ \gamma_k}\ b_k\ooo{}\al{in}(t)+y_k\ooo(t)
\end{eqnarray}
where we have introduced the photon-mediated phonon-phonon interaction matrix
\begin{eqnarray}
\FF_{k,k'}=
\sum_{j=1}^M \GG_{j,k}^*\ \GG_{j,k'} \pq{\wt\chi_j(\omega_0)-\wt\chi_j(-\omega_0)^*
}\ ,
\nn\\
\end{eqnarray}
the corresponding photon-induced mechanical noise operator
\begin{eqnarray}\label{ykooo}
y_k\ooo(t)&=&-\ii\ \ee^{\ii\ \omega_0\ t}\sum_{j=1}^M\pq{
\GG_{j,k}^*\ a_j(t)\Bigl|_{\GG=0}
+
\GG_{j,k}\ a_j\da(t)\Bigl|_{\GG=0}
}
\\&=&
-\frac{\ii}{\sqrt{2\,\pi}}
\int_{-\infty}^\infty\, \dd\,\omega\ 
\ee^{\ii\pt{\omega_0-\omega}\ t}
\nn\\&&\hspace{-1cm}\times
\sum_{j=1}^M\sqrt{2\,\kappa_j}\pq{
\GG_{j,k}^*\ \wt\chi_j(\omega)\ \wt a_j\al{in}(\omega)
+
\GG_{j,k}\ \wt\chi_j(-\omega)^*\ \wt a_j\al{in}{}\da(\omega)
}
\nn
\end{eqnarray}
(note that here we use the notation $\pq{x(\omega)}\da=x\da(-\omega)$) and the slowly varying zero-th optical mode $a_0\ooo(t)=a_0(t)\ \ee^{\ii\ \epsilon_{L0}\ t}$ which, in turn, fulfills the equation
\begin{eqnarray}
\dot a_0\ooo&=&-\pq{\kappa_0+\ii\pt{\Delta_0-\epsilon_{L0}
}}\ a_0\ooo-\ii\sum_{k=1}^N\,\GG_{0,k}\pt{
b_k\ooo+\ee^{2\,\ii\ \omega_0\ t}\ b_k\ooo{\da}}
\nn\\&&+\sqrt{2\,\kappa_0}\ a_0\ooo{}\al{in}(t)\ .
\end{eqnarray}
In Eq.~\rp{QLE01} of the main text we also neglected the non resonant terms from this equation, and we have used the definition
\begin{eqnarray}
\YY_{k,k'}&=&\gamma_k\ \delta_{k,k'}+\frac{\FF_{k,k'}+\FF_{k',k}^*}{2}
\end{eqnarray}
for the dissipative part of the dynamics [which is equal to Eq.~\rp{YY} of the main text], and
\begin{eqnarray}
\WW_{k,k'}&=&\delta\omega_k\ \delta_{k,k'}-\ii\frac{\FF_{k,k'}-\FF_{k',k}^*}{2}\ ,
\end{eqnarray}
for the Hamiltonian part [which is equal to Eq.~\rp{WW} of the main text].

Finally we study the correlation of the photon induced noise operators~\rp{ykooo} and we show that they can be approximated by Eq.~\rp{avyy}.

Specifically we note that the correlation function $\av{y_k\ooo(t)\ y_{k'}\ooo{}\da(t+\tau)}$ should decay over the fast time scale of the cavity dynamics. Hence, if $o(\tau)$ is a generic slow quantity we can approximate
\begin{eqnarray}
\int\dd\tau\ o(\tau)\av{y_k\ooo(t)\ y_{k'}\ooo{}\da(t+\tau)}&\simeq&
\frac{o(t)}{2\,\pi}\ 
\int\dd\tau
\\&&\hspace{-4.3cm}\times
\int_{-\infty}^\infty\, \dd\,\omega\ 
\int_{-\infty}^\infty\, \dd\,\omega'\ 
\ee^{\ii\pt{\omega_0-\omega}\ t}
\ee^{-\ii\pt{\omega_0+\omega'}\ (t+\tau)}
\sum_{j,j'=1}^M 
\nn\\&&\hspace{-4.3cm}\times\
2\sqrt{\kappa_j\,\kappa_{j'}}\
\GG_{j,k}^*\ 
\GG_{j',k'}\
\wt\chi_j(\omega)\
\wt\chi_{j'}(-\omega')^*\
\av{
\wt a_j\al{in}(\omega)\
\wt a_{j'}\al{in}{}\da(\omega')
}
\nn
\end{eqnarray}
and using $\av{
\wt a_j\al{in}(\omega)\
\wt a_{j'}\al{in}{}\da(\omega')
}=\delta_{j,j'}\ \delta(\omega+\omega')$ we find
\begin{eqnarray}
\int\dd\tau\ o(\tau)\av{y_k\ooo(t)\ y_{k'}\ooo{}\da(t+\tau)}&\simeq&
\frac{o(t)}{2\,\pi}\ 
\int\dd\tau\
\\&&\hspace{-3cm}\times
\int_{-\infty}^\infty\, \dd\,\omega\ 
\ee^{-\ii\pt{\omega_0-\omega}\ \tau}
\sum_{j=1}^M2\,\kappa_j\
\GG_{j,k}^*\ 
\GG_{j,k'}\
\abs{\wt\chi_j(\omega)}^2
\nn\\&&\hspace{-3cm}=
o(t)\ \sum_{j=1}^M2\,\kappa_j\
\GG_{j,k}^*\ 
\GG_{j,k'}\
\abs{\wt\chi_j(\omega_0)}^2\ ,
\end{eqnarray}
indicating that we can approximate
\begin{eqnarray}
\av{y_k\ooo(t)\ y_{k'}\ooo{}\da(t')}&\simeq&2\,\delta(t-t')
\sum_{j=1}^M\kappa_j\
\GG_{j,k}^*\ 
\GG_{j,k'}\
\abs{\wt\chi_j(\omega_0)}^2\ .
\nn\\
\end{eqnarray}
Similarly we find
\begin{eqnarray}
\av{y_k\ooo{}\da(t)\ y_{k'}\ooo(t')}&\simeq&2\,\delta(t-t')
\sum_{j=1}^M\kappa_j\
\GG_{j,k}^*\ 
\GG_{j,k'}\
\abs{\wt\chi_j(-\omega_0)}^2\ .
\nn\\
\end{eqnarray}
The correlations $\av{y_k\ooo(t)\ y_{k'}\ooo(t')}$ and $\av{y_k\ooo{}\da(t)\ y_{k'}\ooo{}\da(t')}$, instead, include fast rotating terms, at frequencies $\pm2\,\omega_0$, that mediates to zero their effect on the mechanical dynamics. And thus, we find the results of Eq.~\rp{avyy}.

\section{Considerations on the preparation of cluster states defined on generic graphs}\label{App:general-graphs}

In general, given a cluster state determined by an adjacency matrix $\AAA$, it can be generated by a multimode squeezing transformation~\rp{U}, that is defined in terms of a matrix $\ZZ$~\rp{ZZ}-\rp{ZZ0}. We have found that Eq.~\rp{VVPhiVV} expresses the relation between the matrix $\ZZ$ and the matrices $\VV$~\rp{VV} and $\Phi$, that also determine the Hamiltonian~\rp{H} according to Eqs.~\rp{JJ1}, \rp{WWJ} and \rp{vg0J}. 
Eq.~\rp{VVPhiVV} entails that
\begin{eqnarray}\label{VVPhi}
\VV\ \Phi&=&\sqrt{-\ii\,\ZZ}\ \OO
\nn\\&=&
\Theta\ \sqrt{-\ii\,\ZZ_0}\ \OO_0\ ,
\end{eqnarray}
where $\OO$ and $\OO_0$ are generic orthogonal matrices related by
\begin{eqnarray}\label{OOZ0}
\OO=\OO_\ZZ\ \OO_0\ ,
\end{eqnarray}
with $\OO_\ZZ$ the real orthogonal matrix
\begin{eqnarray}\label{OOZ}
\OO_\ZZ=\pt{\sqrt{-\ii\,\ZZ}}^*\ \Theta\ \sqrt{-\ii\,\ZZ_0}\ .
\end{eqnarray}
The fact that $\OO_Z$ is real orthogonal can be easily proved by showing that $\OO_\ZZ^T\ \OO_\ZZ=\id$ and $\OO_\ZZ\da\ \OO_\ZZ=\id$ (note that both $\sqrt{-\ii\,\ZZ}$ and $\sqrt{-\ii\,\ZZ_0}$ are unitary symmetric).

As discussed in Sec.\ref{Sec:constraints}, in our case, the matrices $\VV$ and $\Phi$ have to fulfill also the relations in Eqs.~\rp{argV} and \rp{VVJ}. Using Eq.~\rp{VVPhi} in Eq.~\rp{argV}, we find that the phases in $\Theta$ are fixed (up to multiples of $\pi$). Specifically, for any $k$,  we can choose $\theta_k$ such that
$\arg\pg{
\pq{
\Theta\ \sqrt{-\ii\,\ZZ_0}\ \OO_0\ \Phi^*}_{k,1}}=\phi_x+n_k\,\pi$
for any specific value $\phi_x$ (the same for all $k$) and an integer $n_k$, that is  
\begin{eqnarray}\label{argV2}
\theta_k=\arg\pg{
\pq{
\sqrt{-\ii\,\ZZ_0}\ \OO_0}_{k,1}}-\frac{\varphi_1}{2}-\phi_x-n_k\,\pi\ .
\end{eqnarray}
Moreover, using Eq.~\rp{VVPhi}-\rp{OOZ}, the relation~\rp{VVJ} can be expressed as
\begin{eqnarray}\label{VVJ2}
\ovl\JJ\al{S}\ \XX + \XX\ \ovl\JJ\al{S}=0\ ,
\end{eqnarray}
with the symmetric unitary matrix
\begin{eqnarray}\label{XX}
\XX&=&\ii\,\OO^T\ \ZZ\ \OO
\nn\\&=&
-\ii\,\OO_0^T\ \sqrt{-\ii\,\ZZ_0}\ \Theta^2
\ \sqrt{-\ii\,\ZZ_0}\ \OO_0\ .
\end{eqnarray}
In this way we obtain the matrix $\VV$ by finding the orthogonal matrix $\OO$ (or $\OO_0$) that fulfill Eqs.~\rp{argV2} and \rp{VVJ2}. 

We note that
\begin{itemize}
\item[(i)]
The values of the squeezing phases $\Phi$ are not fixed by the condition~\rp{argV2}, and~\rp{VVJ2} meaning that we can select any value for $\Phi$ (in particular we can always set $\Phi=\id$); 

\item[(ii)]
The matrix $\OO$ (and $\OO_0$), depends both on the specific choice of the values of the coefficients $J_k$ that constitute the matrix $\ovl\JJ\al{S}$ [see Eq.~\rp{VVJ2}] and on the values of the phases $\theta_k$ that constitute the matrix $\Theta$ [see Eqs.~\rp{argV2} and \rp{VVJ2}];

\item[(iii)]
The steady state of the model of Ref.~\cite{zippilli2021} is independent from the values of the entries of both $\ovl\JJ\al{S}$ and $\Theta$. In fact, on the one hand, $\ovl\JJ\al{S}$ determines the auxiliary chain Hamiltonian in Ref.~\cite{zippilli2021} and the results of Ref.~\cite{zippilli2021} are independent from the specific values of the interactions strengths in $\ovl\JJ\al{S}$; on the other hand the phases in $\Theta$ correspond to local rotations of the modes that constitute the cluster state, so that its global entanglement properties are independent from these phases, and any cluster state with different $\Theta$ (and equal $\ZZ_0$) can be considered equivalent. As a consequence, in the present work, the matrices $\ovl\JJ\al{S}$ and $\Theta$  can be adjusted to determine the matrix $\OO_0$ corresponding to a given cluster state;

\item[(iv)]
Eq.~\rp{argV2} fix the matrix $\Theta$, given $\OO_0$. So, we can express $\Theta$ as the function of $\OO_0$ that corresponds to Eq.~\rp{argV2},
\begin{eqnarray}
\Theta\equiv\ovl\Theta\pq{\OO_0}\ ,
\end{eqnarray}
and Eq.~\rp{VVJ2} can be expressed as
\begin{eqnarray}\label{VVJ3}
&&\ovl\JJ\al{S}\ \OO_0^T\ \sqrt{-\ii\,\ZZ_0}\ \ovl\Theta\pq{\OO_0}^2
\ \sqrt{-\ii\,\ZZ_0}\ \OO_0 
\nn\\&&
+\ \OO_0^T\ \sqrt{-\ii\,\ZZ_0}\ \ovl\Theta\pq{\OO_0}^2
\ \sqrt{-\ii\,\ZZ_0}\ \OO_0\ \ovl\JJ\al{S}=0\ ;
\end{eqnarray}

\item[(v)]
Additionally, $\OO_0$ have to be orthogonal
\begin{eqnarray}\label{othOO}
\OO_0\ \OO_0^T=\id\ ;
\end{eqnarray}

\item[(vi)] 
Finally, the matrix $\VV$, which determines the system Hamiltonian that can be realized with our optomechanial system, for any given cluster state is 
\begin{eqnarray}
\VV&=&
\ovl\Theta\pq{\OO_0} \sqrt{-\ii\,\ZZ_0}\ \OO_0\ ,
\end{eqnarray}
where $\OO_0$ is solution of Eqs.~\rp{VVJ3} and\rp{othOO}.

\end{itemize}

We have not found a general solution for this problem. However we have identified simple solutions for the cases of rectangular graphs with at least one side of the graph made by an odd number of nodes (see Sec.~\ref{Sec:rectgraph}). It should be possible to find more general solutions by approaching this problem numerically.

\section{Steady-state solution and covariance matrix}
\label{App:Standard method}

To evaluate the results of Sec.~\ref{Sec:results} we have computed the steady state correlation matrix $\CC$, corresponding to the full model~\rp{QLE00}. The correlation matrix $\CC$ is the $2(N+M+1)\times2(N+M+1)$ matrix with elements $\CC_{\ell,\ell'}=\av{\va_\ell\ \va_{\ell'}}$, where the symbol $\va$ indicates the vector of operators
\begin{eqnarray}
\va=\pt{a_0,...a_M,b_1,...b_N,a_0\da,...a_M\da,b_1\da,...b_N\da}\ .
\end{eqnarray} 
Using Eq.~\rp{QLE00} it is straightforward to determine the corresponding equation for $\CC$. It can be expressed in matrix form as 
\begin{eqnarray}
\dot\CC(t)=\MM\ \CC(t)+\CC(t)\ \MM^T+\NN(t)\ ,
\end{eqnarray}
in terms of drift matrix 
\begin{eqnarray}
\MM=-\pt{\mmat{cccc}{
\KK          & \ii\,\GG    & 0           & \ii\,\GG   \\
\ii\,\GG\da  & \YY         & \ii\,\GG^T  & 0           \\
0            & -\ii\,\GG^* & \KK^*       & -\ii\,\GG^*  \\
-\ii\,\GG\da & 0           & -\ii\,\GG^T & \YY^*
}}\ ,
\end{eqnarray} 
where $\KK\in\mathds{C}^{(M+1)\times (M+1)}$ is diagonal with entries $\KK_{j,j}=\kappa_j+\ii\,\Delta_j$, $\YY\in\mathds{C}^{N\times N}$ is diagonal with entries $\YY_{k,k}=\gamma_k+\ii\,\omega_k$, and $\GG\in\mathds{C}^{(M+1)\times N}$ is the coupling matrix defined in Eq.~\rp{GGjk}. Finally, $\NN(t)$ is the diffusion matrix, which is time dependent because of the correlation functions of the squeezed bath~\rp{corrSqueezBath}~\cite{asjad2016a}.
It can be decomposed as
\begin{eqnarray}
\NN(t)=\NN_0+\NN_{n}
+\NN_{m}\al{-}\ \ee^{-2\,\ii\,\epsilon_{L0} t}
+\NN_{m}\al{+}\ \ee^{2\,\ii\,\epsilon_{L0} t}
\end{eqnarray}
where
\begin{eqnarray}
\NN_0=\pt{\mmat{cccc}{
0 & 0                           & \KK+\KK^* & 0                                    \\
0 & 0                           & 0         & \pt{\YY+\YY^*}\pt{\id+\ovl\NN_{T,b}} \\
0 & 0                           & 0         & 0                                    \\
0 & \pt{\YY+\YY^*}\ovl\NN_{T,b} & 0         & 0
}}\ ,
\end{eqnarray} 
with the zeros indicating null matrices and $\ovl\NN_{T,b}\in\mathds{C}^{N\times N}$ diagonal with entries $\pg{\ovl\NN_{T,b}}_{k,k}=\ovl n_{T,k}$; $\NN_{n}$ is a matrix with only two non-zero elements [those with indices $\pt{1,N+M+2}$ and $\pt{N+M+2,1}$] that are equal to
\begin{eqnarray}
\pg{\NN_{n}}_{1,N+M+2}=\pg{\NN_{n}}_{N+M+2,1}=2\,\kappa_0\ n_s\ ;
\end{eqnarray}
and finally $\NN_{m}\al{-}$ and $\NN_{m}\al{+}$ have both a single non-zero element
\begin{eqnarray}
\pg{\NN_{m}\al{-}}_{1,1}=\pg{\NN_{m}\al{+}}_{N+M+2,N+M+2}^*=2\,\kappa_0\ m_s\ .
\end{eqnarray}
When the system is stable, that is when the real parts of all the eigenvalue of $\MM$ are negative, then the steady state solution can be formally expressed as~\cite{asjad2016a}
\begin{eqnarray}\label{Cst}
\CC_{\rm st}&=&-\LL^{-1}\pt{\NN_0+\NN_n}
\nn\\&&
-\pt{\LL+2\,\ii\ \epsilon_{L0}}^{-1}\NN_m\al{-}\ \ee^{-2\,\ii\,\epsilon_{L0} t}
\nn\\&&
-\pt{\LL-2\,\ii\ \epsilon_{L0}}^{-1}\NN_m\al{+}\ \ee^{2\,\ii\,\epsilon_{L0} t}\ ,
\end{eqnarray}
where $\LL$ is the linear operator defied by $\LL\,\CC=\MM\,\CC+\CC\,\MM^T$.
The reduced correlation matrix for the mechanical modes, can be written 
as
\begin{eqnarray}
\CC_{\rm st}\al{b}=\SSS\ \CC_{\rm st}\ \SSS^T\,
\end{eqnarray}
with the $2(N+M+1)\times2N$ matrix
\begin{eqnarray}
\SSS=\pt{\mmat{cccc}{
0&\id_N&0&0\\
0&0    &0&\id_N
}}\ ,
\end{eqnarray}
where the zeros indicate null matrices and $\id_N$ is the $N\times N$ identity matrix.
Here we are interested in the slowly varying mechanical variable~\rp{slowOp}, and their correlation matrix is given by 
\begin{eqnarray}\label{Cstbooo}
\CC_{\rm st}\al{b}{}\ooo
=\matt{
\ee^{\ii\,\id_N\,\omega_0\,t}&0
}{
0&\ee^{-\ii\,\id_N\,\omega_0\,t}
}\ 
\CC_{\rm st}\al{b}\ 
\matt{
\ee^{\ii\,\id_N\,\omega_0\,t}&0
}{
0&\ee^{-\ii\,\id_N\,\omega_0\,t}
}\ .
\nn\\
\end{eqnarray}

It is useful to also introduce the covariance matrices in the canonical quadrature basis, that is the symmetric matrix of correlation for the quadrature operators $x_k=b_k\ooo+b_k\ooo{}\da$ and $p_k=-\ii\,b_k\ooo+\ii\,b_k\ooo{}\da$. Introducing the vector of operators 
\begin{eqnarray}\label{vx}
\vx=\pt{x_1,...x_N,p_1,...p_N}\ ,
\end{eqnarray}  
the elements of the covariance matrix are $\EE_{k,k'}\al{b}=\pt{\av{\vx_k\ \vx_{k'}}+\av{\vx_{k'}\ \vx_k}}/2$, and the steady state mechanical covariance matrix can be expressed, using the $2N\times2N$ matrix
\begin{equation}
\RR=\pt{
\mmat{cc}{
\id &\id\\
-\ii\,\id &\ii\, \id
}}
\end{equation}
as
\begin{eqnarray}\label{ststcovmat}
\EE_{\rm st}\al{b}=\RR\ \frac{\CC_{\rm st}\al{b}{}\ooo+{\CC_{\rm st}\al{b}{}\ooo}^T}{2}\ \RR^T\ .
\end{eqnarray}

In general, the steady state is time dependent [see Eqs.~\rp{Cst} and \rp{Cstbooo}] and exhibits residual oscillations that are due to non-resonant blue-sideband transitions~\cite{asjad2016a}. In the limit studied here $\epsilon_{L0}=\omega_0$ (such that the zero-th optical mode is resonant with the central frequency of the squeezed reservoir) and small $\kappa_j$, this effect is very small so that deviations in the results at different times are very small (see Ref.~\cite{asjad2016a} for a detailed discussion of this effect). The results reported in Sec.~\ref{Sec:results} are evaluated by setting $t=0$ in Eqs.~\rp{Cst} and \rp{Cstbooo}.

\section{Fidelity and variance of the nullifiers}
\label{App:fidelity-nullifiers}

In Sec.~\ref{Sec:results} we reported results for the steady state fidelity and for the variance of the nullifiers. They are evaluated using the steady state covariance matrix $\EE_{\rm st}\al{b}$~\rp{ststcovmat} as shown below. 

\subsection{Fidelity}

The fidelity measures how close two states are. The fidelity between the steady state of our system and the target state, can be expressed in terms of the covariance matrices as~\cite{spedalieri2013}
\begin{eqnarray}\label{F}
F=\frac{2^N}{\sqrt{\EE_{\rm st}\al{b}+\EE_{\rm target}\al{b}}}\ ,
\end{eqnarray} 
where the target covariance matrix $\EE_{\rm target}\al{b}$ is evaluated as follows.
The elements of $\EE_{\rm target}\al{b}$ are 
\begin{eqnarray}
\pg{\EE_{\rm target}\al{b}}_{k,k'}=
\br{\Psi_{\rm cluster}}
\frac{\vx_k\ \vx_{k'}+\vx_{k'}\ \vx_k}{2}
\ke{\Psi_{\rm cluster}}
\end{eqnarray}
with $\ke{\Psi_{\rm cluster}}$ given by Eq.~\rp{Psicluster}.
Using Eqs.~\rp{Psicluster} and \rp{UdacU}, and the equivalence between operators given by Eq.~\rp{cb}], one can show that the target covariance matrix can be expressed 
in matrix form as
\begin{eqnarray}
\EE_{\rm target}\al{b}=\RR\ \BBB\ \pt{\mmat{cc}{
0&\id_N\\
\id_N&0
}} \BBB^T\ \RR^T\ ,
\end{eqnarray}
where
\begin{eqnarray}
\BBB=\pt{\mmat{cc}{
\cosh(z)\ \VV & \sinh(z)\ \VV\ \Phi^2\\
\sinh(z)\ \VV^*\ \Phi^*{}^2 &\cosh(z)\ \VV^*
}}\ .
\end{eqnarray}

\subsection{Variance of the nullifiers}
\label{App:avX}

The other quantity that we study in Sec.~\ref{Sec:results} is the variance of the nullifiers~\rp{nulli}.

The variance of the nullifiers $\av{X_k^2}$, evaluated over the steady state, can be expressed in terms of the covariance matrix~\rp{ststcovmat}, by rewriting the 
nullifiers as $X_k=\pg{\QQ\ \vx}_k$ in terms of the $N\times 2N$ block matrix
\begin{eqnarray}
\QQ=\pt{\Theta\al{s}-\AAA\,\Theta\al{c},\Theta\al{c}+\AAA\,\Theta\al{s}}\ ,
\end{eqnarray}
where $\Theta\al{c}=\pt{\Theta^*+\Theta}/2$ and $\Theta\al{s}=\pt{\Theta^*-\Theta}/2\ii$.
Using this expression we find that the steady state variance of the nullifiers is given by the diagonal elements of the matrix $\QQ\ \EE_{\rm st}\al{b}\ \QQ^T$, i.e.
\begin{eqnarray}\label{avX}
\av{X_k^2}_{\rm st}=\pg{\QQ\ \EE_{\rm st}\al{b}\ \QQ^T}_{k,k}\ .
\end{eqnarray}

\input{OMClusters_SqueezRes.bbl}


\end{document}

%% file: OMClusters_SqueezRes.bbl
%